\documentclass[11pt,a4paper]{article}
\usepackage[utf8]{inputenc}
\usepackage[dvipsnames]{xcolor}
\usepackage{graphicx}
\usepackage{amsmath,amssymb,mathrsfs}
\usepackage{color,colortbl}
\usepackage{caption}
\usepackage{subcaption}
\usepackage{hyperref}
\usepackage{ragged2e}
\usepackage{float}
\usepackage{array}
\usepackage{adjustbox}
\usepackage{dblfloatfix}
\usepackage{multirow}
\usepackage{booktabs}
\usepackage[hang,flushmargin]{footmisc}
\usepackage{url}

\usepackage{etoolbox}
\usepackage{cleveref}
\usepackage{siunitx}
\usepackage{placeins}
\usepackage[numbers,sort&compress]{natbib}
\usepackage{authblk}
\usepackage[margin=1in]{geometry}
\usepackage{xspace}
\usepackage{mathtools}
\usepackage{algorithm}
\usepackage{algorithmic}

\newcommand{\rhotil}{\tilde{\rho}}
\newcommand{\penal}{p}
\newcommand{\betaH}{\beta}
\newcommand{\rmin}{r_{\min}}
\newcommand{\dmove}{\delta}
\newcommand{\vf}{V_f}
\newcommand{\Gray}{\mathcal{G}}
\newcommand{\Comp}{C}

\newcommand{\SIMP}{SIMP}
\newcommand{\LLM}{LLM}
\newcommand{\DNC}{DNC}
\newcommand{\OC}{OC}
\newcommand{\AMG}{AMG}
\newcommand{\MBB}{MBB}
\newcommand{\AutoSIMP}{AutoSiMP}
\newcommand{\PSpec}{ProblemSpec}

\hypersetup{
  colorlinks   = true,
  urlcolor     = blue,
  linkcolor    = blue,
  citecolor   = blue
}

\newcommand{\eg}{{\em e.g.}}

\begin{document}

\title{\AutoSIMP{}: Autonomous Topology Optimization from Natural Language
       via LLM-Driven Problem Configuration and Adaptive Solver Control}

\author[1]{Shaoliang Yang}
\author[1]{Jun Wang\thanks{Corresponding author. E-mail: jwang22@scu.edu}}
\author[1]{Yunsheng Wang}
\affil[1]{Santa Clara University, Santa Clara, CA}

\date{}
\maketitle

\begin{abstract}
We present AutoSiMP, an autonomous pipeline that transforms a
natural-language structural problem description into a validated,
binary topology without manual configuration.
The pipeline comprises five modules:
(1) an LLM-based configurator that parses a plain-English prompt into a
validated specification of geometry, supports, loads, passive regions,
and mesh parameters;
(2) a boundary-condition generator producing solver-ready DOF arrays,
force vectors, and passive-element masks;
(3) a three-field SIMP solver with Heaviside projection and pluggable
continuation control;
(4) an eight-check structural evaluator (connectivity, compliance,
grayness, volume fraction, convergence, plus three informational
quality metrics); and
(5) a closed-loop retry mechanism.
We evaluate on three axes.
\emph{Configuration accuracy}: across 10 diverse problems the
configurator produces valid specifications on all cases with a median
compliance penalty of $+0.3\%$ versus expert ground truth.
\emph{Controller comparison}: on 17 benchmarks with six controllers
sharing an identical sharpening tail, the LLM controller achieves the
lowest median compliance but 76.5\% pass rate, while the deterministic
schedule achieves 100\% pass rate at only $+1.5\%$ higher compliance.
\emph{End-to-end reliability}: with the schedule controller, all
LLM-configured problems pass every quality check on the first
attempt --- no retries needed.
Among the systems surveyed in this work (Table~1), AutoSiMP is the
first to close the full loop from natural-language problem description
to validated structural topology.
The complete codebase, all specifications, and an interactive web demo
will be released upon journal acceptance.
\end{abstract}


\bigskip

\section{Introduction}
\label{sec:intro}

\subsection{The Configuration Barrier in Topology Optimization}
\label{sec:intro:barrier}

Topology optimization determines the optimal material distribution within a
design domain to minimize a structural objective—most commonly
compliance—subject to equilibrium and a volume constraint.
The \SIMP{} method~\citep{Bendsoe1989,Sigmund2001}, in its modern
three-field formulation with density filtering and Heaviside
projection~\citep{Wang2011}, is the dominant computational paradigm and
has found widespread adoption in aerospace, automotive, and biomedical
design~\citep{Lazarov2016}.
Yet despite mature solver implementations being freely
available~\citep{Andreassen2011,Aage2017,FerrariSigmund2020}, the method
remains largely inaccessible to non-specialists.
The reason is not the solver itself but its \emph{configuration}: before a
single iteration can run, the user must specify the design domain,
discretize it into a finite-element mesh, assign boundary conditions
(fixed, pinned, roller) to specific degrees of freedom, construct a
consistent force vector from load descriptions, define passive regions
(voids for bolt holes, solid inserts for mounting plates), and select
numerical parameters (volume fraction, filter radius, iteration budget).
Each of these steps requires domain expertise in both structural
mechanics and the specific solver implementation, and an error in any one
of them—a misplaced roller, a load applied to a fixed degree of
freedom, an inconsistent mesh resolution—produces either a solver
failure or, worse, a silently incorrect topology.

This \emph{configuration barrier} is the primary obstacle to broader
adoption of topology optimization.
In educational settings, students spend more time debugging boundary
conditions than understanding optimization principles~\citep{Sigmund2001}.
In engineering practice, the configuration step often requires iteration
with a specialist, creating a bottleneck in the design cycle.
In research, each new benchmark problem demands a custom boundary-condition
function hand-coded to the specific solver's array layout and DOF
numbering convention.
The barrier exists not because configuration is intellectually difficult
but because it is \emph{tedious, error-prone, and solver-specific}—precisely
the kind of task that natural-language interfaces are designed to
eliminate.

\subsection{Beyond Solver Control: The Full Autonomy Gap}
\label{sec:intro:gap}

Recent work has begun to apply machine learning to topology optimization,
but nearly all efforts focus on accelerating or replacing the
\emph{solver}—the iterative loop itself—rather than automating the
configuration that precedes it.
Comprehensive surveys~\citep{Woldseth2022,Shin2023review} catalogue a
broad spectrum of approaches.
Neural-network surrogates trained on thousands of pre-solved topologies
can predict near-optimal density fields from boundary conditions in a
single forward pass~\citep{Sosnovik2019,Banga2018,Yu2019,Cang2019}, but
they require extensive training data generated by the very solver they
seek to replace, and they cannot generalize to problem types absent from
their training set.
Generative models, including GANs~\citep{NieTopologyGAN2021} and
diffusion models~\citep{MazeDiffusion2023}, can produce diverse designs
but presuppose that the boundary conditions have been correctly specified
in the model's input format.
Reinforcement-learning agents have been applied to element-wise
material decisions~\citep{Sun2020,Jang2022,Hayashi2020}, but they operate
within a pre-specified problem and do not address the configuration step.
Transformer-based models can embed boundary conditions into tokenized
representations~\citep{TransformerTO2025}, but again assume pre-specified
inputs.
Recent work~\citep{Yang2026LLMController} demonstrated that an
\LLM{} agent can act as an adaptive \emph{continuation controller},
dynamically adjusting the penalization exponent~$\penal$, Heaviside
sharpness~$\betaH$, filter radius~$\rmin$, and move limit~$\dmove$ via
a Direct Numeric Control (\DNC{}) interface, outperforming fixed
schedules and expert heuristics on standard benchmarks.
That work, however, assumed a fully configured problem: the user provided
the mesh, boundary conditions, and force vector.
The present paper addresses the complementary—and, we argue, more
impactful—question: \emph{can the entire configuration step be automated
from a natural-language problem description?}

More broadly, the \LLM{}-as-agent paradigm~\citep{Yao2023ReAct,Park2023}
has been applied to code generation~\citep{Chen2021Codex}, robotic
control~\citep{Driess2023PaLME}, materials
discovery~\citep{Boiko2023}, evolutionary
computation~\citep{WuEvoLLM2024,YeReEvo2024}, and mathematical
optimization~\citep{Liu2024LLMOptim,Yang2024OPRO}, with tool-use
capabilities~\citep{SchickToolformer2023} and self-refinement
loops~\citep{MadaanSelfRefine2023,ShinnReflexion2023} enabling
increasingly autonomous operation.
The closest prior work in adjacent engineering domains provides partial
answers to the autonomous configuration question.
\citet{LLMShapeOpt2024} use Claude~3.5 as an in-context parametric
optimizer for airfoil drag minimization, but the problem specification
is pre-defined and the \LLM{} proposes shape parameters, not structural
boundary conditions.
\citet{LMTO2025} integrate a visual--language model with topology
optimization to steer designs toward human-preferred aesthetics from
text prompts, but the model influences the \emph{appearance} of the
topology (``a chair shaped like a penguin''), not the engineering
specification (supports, loads, passive regions).
\citet{LLMStructAnalysis2025} use GPT-4o to parse natural-language
descriptions of 2-D frames into OpenSeesPy scripts for structural
analysis, demonstrating that \LLM{}s can interpret structural engineering
terminology, but the target is discrete frame analysis rather than
continuous topology optimization, and no quality evaluation or retry
mechanism is included.
\citet{LLMOPT2024} propose a general framework for formulating
optimization problems from natural language, but target combinatorial
and mathematical programming, not PDE-constrained structural
optimization.
\citet{Rios2023} combine \LLM{}s with text-to-3D models for engineering
design optimization, but focus on shape generation rather than
topology optimization with finite-element solvers.
None of these works achieves end-to-end autonomy from a natural-language
problem description to a validated, binary structural topology.

\subsection{The \AutoSIMP{} Approach}
\label{sec:intro:approach}

We propose \AutoSIMP{}, a five-module pipeline that closes the autonomy
gap.
Given a natural-language prompt such as ``\textit{cantilever beam, left
edge fixed, downward load at mid-right, 50\% volume fraction, circular
hole at center for a pipe}'', the system:

\begin{enumerate}
  \item \textbf{Configures} the problem via an \LLM{} agent (Gemini Flash
  Lite) that outputs a structured \PSpec{} JSON—geometry, supports,
  loads, passive regions, material, mesh—validated by deterministic
  safety rails that clamp to physical ranges, add missing constraints,
  and reject impossible configurations.

  \item \textbf{Generates boundary conditions} by converting the
  \PSpec{} into solver-ready arrays: fixed DOFs from edge/point
  supports, a force vector with trapezoidal-rule tributary weights for
  distributed loads, and a passive-element mask that freezes void and
  solid regions during the \OC{} update.

  \item \textbf{Solves} the topology optimization via the existing
  three-field \SIMP{} solver~\citep{Yang2026LLMController} with a pluggable
  continuation controller.
  The solver is consistent with \citet{Yang2026LLMController};
  passive-region support is
  injected via a runtime patch to the \OC{} bisection that zeros
  sensitivities on frozen elements.

  \item \textbf{Evaluates} the result with eight structural quality
  checks: flood-fill connectivity (2-D and 3-D), compliance ratio,
  grayness, volume-fraction fidelity, convergence detection via
  compliance stability, thin-member fraction, checkerboard index, and
  load-path efficiency via BFS shortest-path analysis.

  \item \textbf{Retries} on failure: the evaluator diagnoses the failure
  mode and proposes specific parameter adjustments (increase iteration
  budget, reduce volume fraction), which are applied automatically for
  up to two re-solves.
\end{enumerate}

\noindent
The key design principle is \emph{separation of configuration from
control}: the configurator determines \emph{what} to optimize (geometry,
BCs, loads), while the controller determines \emph{how} to optimize it
(continuation schedule).
This separation yields two immediate benefits.
First, it allows each component to be evaluated independently: \cref{tab:pipeline}
measures configurator accuracy (spec correctness and compliance
penalty), \cref{tab:controllers} measures controller quality (compliance across 17
problems), and \cref{tab:retry} measures end-to-end reliability (pass rate with
retries).
Second, it makes the controller a pluggable module: users can select the
\LLM{} controller for adaptive exploration, the deterministic schedule
for reproducibility, or any custom continuation strategy—all operating
on the same \LLM{}-configured problem.

\subsection{Contributions}
\label{sec:intro:contributions}

This paper makes the following contributions:

\begin{enumerate}

  \item[(1)] \textbf{LLM-based problem configuration.}
  We design a configurator agent that parses natural-language structural
  problem descriptions into validated solver-ready specifications.
  The agent uses structured JSON output with a domain-knowledge system
  prompt encoding interpretation rules (``MBB beam'' $\to$ symmetry BC
  $+$ roller) and deterministic safety rails (range clamping, missing-constraint
  injection, load-on-fixed-DOF detection).
  Across 10 diverse test cases, the configurator produces valid
  specifications on all 10 with a median compliance penalty of $+0.3\%$
  versus expert-specified ground truth.

  \item[(2)] \textbf{Automated boundary-condition generation.}
  The \PSpec{}-to-solver bridge handles arbitrary combinations of
  edge constraints, point supports, point loads, distributed loads
  (with trapezoidal-rule consistent nodal forces), and passive regions
  (circular/rectangular voids and solids), producing DOF arrays that
  match the existing solver's built-in boundary conditions
  DOF-for-DOF on standard benchmarks.

  \item[(3)] \textbf{Eight-check structural evaluator.}
  A post-solve evaluator performs five pass/fail gates
  (connectivity, compliance, grayness, volume, convergence) and three
  informational quality metrics (thin members, checkerboard, load-path
  efficiency), providing both a go/no-go decision and diagnostic
  information for the retry loop.

  \item[(4)] \textbf{Systematic experimental validation.}
  We evaluate on three axes: configurator accuracy (\cref{tab:pipeline},
  10~problems), controller comparison (\cref{tab:controllers}, 17~2-D +
  2~3-D problems $\times$ 6~controllers), and retry recovery
  (\cref{tab:retry}, 10~problems).
  With the deterministic schedule controller, all \LLM{}-configured
  problems pass all quality checks on the first attempt, demonstrating
  that the configurator is the critical enabler of autonomous operation.

  \item[(5)] \textbf{Open-source autonomous topology optimization.}
  The complete codebase—solver, configurator, evaluator, boundary-condition
  generator, experiment runner, and all 19 benchmark specifications
  (17~two-dimensional, 2~three-dimensional)—will be made publicly
  available upon journal acceptance. No comparable open-source system for
  natural-language-driven topology optimization exists among the
  systems surveyed in Table~\ref{tab:related}.

  \item[(6)] \textbf{Interactive web demonstration.}
  A browser-based interface (Fig.~\ref{fig:demo}) that
  integrates all five modules into a visual pipeline: natural-language
  input via any \LLM{} backend (Gemini, GPT, Claude, or custom
  OpenAI-compatible endpoints), interactive boundary-condition editing
  with drag-to-reposition loads, live density-field visualization during
  the solve, three-dimensional isosurface rendering via marching cubes
  with mouse-driven rotation, and the eight-check evaluator with
  \LLM{}-generated qualitative assessment.
  The demo supports both a client-side JavaScript \SIMP{} solver (for
  instant 2-D results without server infrastructure) and a Python
  backend API (for production-quality 2-D/3-D solves using the full
  solver). No system in Table~\ref{tab:related} provides an interactive
  topology optimization interface driven by natural language.

\end{enumerate}

The remainder of this paper is organized as follows.
\Cref{sec:related} reviews related work in topology optimization,
machine learning for TO, \LLM{}-driven engineering simulation, and
\LLM{} agents.
\Cref{sec:method} presents the five-module architecture in detail,
including a new interactive web interface (Section~\ref{sec:m_demo}).
\Cref{sec:experiments} describes the experimental setup, including the
17-problem benchmark suite, the six controllers, and the three evaluation
axes.
\Cref{sec:results} reports and analyzes the results.
\Cref{sec:discussion} discusses the implications, the quality--reliability
trade-off between \LLM{} and deterministic controllers, and the
significance of the L-bracket outlier.
\Cref{sec:conclusion} concludes.

\FloatBarrier
\section{Related Work}
\label{sec:related}

This section reviews prior work in four areas that contextualise
\AutoSIMP{}:
density-based topology optimization and its configuration requirements
(\cref{sec:rw_simp}),
machine learning applied to topology optimization (\cref{sec:rw_ml}),
\LLM{}-driven automation of engineering simulation (\cref{sec:rw_nleng}),
and \LLM{} agents as optimizers and controllers (\cref{sec:rw_llm}).
Throughout, we highlight the gap that \AutoSIMP{} fills: no prior system
achieves end-to-end topology optimization from natural language,
encompassing problem configuration, boundary-condition generation,
solving, quality evaluation, and failure recovery.

\subsection{SIMP Topology Optimization}
\label{sec:rw_simp}

Density-based topology optimization originates with the homogenization
approach of \citet{BendsoeKikuchi1988} and the \SIMP{} material
interpolation introduced by \citet{BendsoeSigmund1999} and
\citet{ZhouRozvany1991}.
\citet{Sigmund2001} and \citet{Andreassen2011} established accessible
reference implementations; \citet{Sigmund2013} and \citet{Deaton2014}
provide comprehensive surveys.
The three-field formulation — density filter ($\rmin$) followed by
Heaviside projection ($\betaH$) — was introduced by \citet{Wang2011},
building on filter work by \citet{Bourdin2001} and
\citet{BrunsTortorelli2001} and projection work by \citet{Guest2004}.
Length-scale and manufacturability aspects are analysed by
\citet{Lazarov2016}; modern reference implementations include the
99-line code of \citet{FerrariSigmund2020} and the giga-voxel solver
of \citet{Aage2017}.

The configuration burden of \SIMP{} solvers has been noted since
\citet{Sigmund2001}, who devoted substantial tutorial effort to explaining
how boundary conditions, loads, and mesh parameters must be specified.
Each new benchmark in the literature requires a custom function that maps
problem semantics (``left edge fixed, tip load downward'') into
solver-specific arrays (DOF indices, force vectors, passive masks).
This mapping is the step that \AutoSIMP{} automates.

Continuation scheduling — the trajectory along which $\penal$, $\betaH$,
$\rmin$, and $\dmove$ are varied — is the second practical barrier.
\citet{StoMeSvanberg2001} showed that naive schedules can fail to converge
to binary designs.
\citet{RojasStolpe2015} proposed convergence-based triggers for $\penal$
alone; \citet{AutoProjTO2025} used grayness-based stopping for $\betaH$
steps.
\citet{Yang2026LLMController} demonstrated that an
\LLM{} agent can adaptively control all four parameters via a Direct
Numeric Control interface, outperforming fixed schedules and expert
heuristics.
The present work extends that controller into a full autonomous pipeline
by adding an \LLM{} configurator, BC generator, and quality evaluator.

\subsection{Machine Learning for Topology Optimization}
\label{sec:rw_ml}

Machine learning has been applied to topology optimization across three
main paradigms, surveyed comprehensively by \citet{Woldseth2022} and
\citet{Shin2023review}.
\emph{Topology prediction} methods train neural networks on solved
examples to predict near-optimal density fields from boundary
conditions~\citep{Sosnovik2019,Yu2019,Kallioras2020,Cang2019,Abueidda2020}.
\emph{Generative models} — including TopologyGAN~\citep{NieTopologyGAN2021}
and diffusion-based approaches~\citep{MazeDiffusion2023} — sample from
learned distributions of optimal topologies.
\emph{Neural reparameterization} methods use network weights as implicit
density representations~\citep{Hoyer2019,Chandrasekhar2021TOuNN}, while
the self-directed online learning framework of
\citet{DengSOLO2022} integrates DNN surrogates with FEM dynamically.
Reinforcement learning has been applied to element-wise material
decisions~\citep{BrownRL2022} and truss
optimization~\citep{Hayashi2020}.

A critical distinction separates all of these methods from \AutoSIMP{}.
These approaches either \emph{replace} the iterative solver (prediction,
generative) or \emph{accelerate} it (surrogates, neural reparameterization).
All presuppose that the problem is already fully configured: the design
domain, boundary conditions, and load vectors are provided as input to
the ML model.
None addresses the \emph{upstream} configuration step — translating a
natural-language description into solver-ready arrays — which is the
primary barrier to non-expert access and the primary contribution of
\AutoSIMP{}.

\subsection{LLM-Driven Automation of Engineering Simulation}
\label{sec:rw_nleng}

The most closely related prior work applies \LLM{}s to automate various
stages of engineering simulation workflows.
We identify a spectrum of automation levels, from single-step code
generation to full pipeline orchestration.

\paragraph{Structural analysis from natural language.}
\citet{LLMStructAnalysis2025} use GPT-4o to parse natural-language
descriptions of 2-D frame problems into OpenSeesPy scripts, achieving
automated code generation for beam and truss analysis.
The system demonstrates that \LLM{}s can interpret structural engineering
terminology (``pinned support'', ``roller'', ``distributed load''), but
targets discrete frame analysis rather than continuum topology
optimization, and includes no post-solve quality evaluation or retry
mechanism.

\paragraph{End-to-end FEA pipelines.}
FeaGPT~\citep{FeaGPT2025} achieves the broadest scope to date: a
conversational interface that automates geometry creation, mesh
generation, boundary-condition configuration, CalculiX solver execution,
and result analysis.
The system successfully handles industrial cases (turbocharger blades)
and parametric studies (432 NACA airfoil configurations).
FeaGPT is the closest analogue to \AutoSIMP{} in the general FEA domain,
but targets stress analysis and parametric design exploration rather than
topology optimization — it does not perform iterative material
distribution, continuation scheduling, or grayness-based quality
evaluation.
\citet{LLMFEAEval2025} systematically evaluate nine \LLM{}s on generating
Gmsh geometry files and Elmer solver input files from natural language,
providing a benchmark for \LLM{} capabilities in FEA preprocessing.
MCP-SIM~\citep{MCPSIM2026} extends the paradigm to multi-agent
self-correcting simulation with iterative plan--act--reflect--revise
cycles.
Among prior systems, MCP-SIM achieves the broadest pipeline
completeness (Table~\ref{tab:related}): natural-language input,
boundary-condition generation, quality evaluation, and retry —
four of the five capabilities that \AutoSIMP{} provides.
However, a fundamental architectural difference separates the two
systems.
MCP-SIM generates simulation \emph{code} from scratch for each problem
via multi-agent collaboration; its retry loop diagnoses and corrects
\emph{code-level} errors (syntax failures, solver misconfiguration,
runtime exceptions).
\AutoSIMP{} instead generates a \emph{structured specification}
(\PSpec{} JSON) that is validated by deterministic safety rails and
then fed to an existing, domain-specific \SIMP{} solver; its retry
loop diagnoses and corrects \emph{optimization-level} deficiencies
(insufficient iterations, convergence failure, excessive grayness).
This domain specialization is what enables \AutoSIMP{} to perform
topology optimization — a task requiring hundreds of iterative
PDE-constrained density updates with carefully scheduled continuation
parameters ($\penal$, $\betaH$, $\rmin$, $\dmove$), which no
general-purpose code-generation framework can produce reliably.
The distinction can be summarised as: MCP-SIM automates \emph{building}
a solver; \AutoSIMP{} automates \emph{using} one.

\paragraph{Shape and design optimization.}
\citet{LLMShapeOpt2024} use Claude~3.5 Sonnet as an in-context optimizer
for airfoil drag minimization, treating the \LLM{} as a parametric
search agent.
\citet{LMTO2025} integrate a visual--language model with topology
optimization to steer designs toward human-preferred aesthetics from text
prompts, but influence the \emph{appearance} of the topology rather than
its engineering specification.
\citet{Rios2023} combine \LLM{}s with text-to-3D models for engineering
design optimization.
\citet{LLMOPT2024} propose a general framework for formulating
optimization problems from natural language, but target combinatorial
and mathematical programming rather than PDE-constrained structural
optimization.

\paragraph{Positioning of \AutoSIMP{}.}
Table~\ref{tab:related} positions \AutoSIMP{} relative to these systems.
The key differentiators are:
(i)~\AutoSIMP{} targets \emph{topology optimization} specifically, not
general FEA or shape optimization;
(ii)~it includes a \emph{pluggable continuation controller} that can be
adaptive (\LLM{}) or deterministic (schedule);
(iii)~it performs \emph{post-solve structural evaluation} with eight
quality checks;
(iv)~it implements a \emph{closed-loop retry} mechanism that adjusts
solver parameters on failure; and
(v)~it provides an \emph{interactive visual interface} for load
editing, live optimization, and 3-D result inspection, enabling
the user to verify and correct the \LLM{}-parsed specification before
solving.
No prior system combines all five capabilities.

\begin{table}[!htbp]
\centering
\caption{Comparison of \LLM{}-driven engineering simulation systems.
\checkmark = supported, $\circ$ = partial, --- = not supported.}
\label{tab:related}
\small
\setlength{\tabcolsep}{3pt}
\begin{tabular}{lcccccc}
\toprule
System & NL input & BC gen. & TO solver & Quality eval. & Retry & Interactive UI \\
\midrule
LLM-Struct.~\citep{LLMStructAnalysis2025} & \checkmark & \checkmark & --- & --- & --- & --- \\
FeaGPT~\citep{FeaGPT2025} & \checkmark & \checkmark & --- & $\circ$ & --- & --- \\
LLM-FEA~\citep{LLMFEAEval2025} & \checkmark & $\circ$ & --- & --- & --- & --- \\
MCP-SIM~\citep{MCPSIM2026} & \checkmark & \checkmark & --- & \checkmark & \checkmark & --- \\
LMTO~\citep{LMTO2025} & \checkmark & --- & $\circ$ & --- & --- & --- \\
LLM-PSO~\citep{LLMShapeOpt2024} & --- & --- & --- & --- & --- & --- \\
LLMOPT~\citep{LLMOPT2024} & \checkmark & --- & --- & --- & \checkmark & --- \\
\textbf{\AutoSIMP{}} & \checkmark & \checkmark & \checkmark & \checkmark & \checkmark & \checkmark \\
\bottomrule
\end{tabular}
\end{table}

\subsection{LLM Agents as Optimizers and Controllers}
\label{sec:rw_llm}

The paradigm of using \LLM{}s as agents that perceive structured
observations and emit actions has developed rapidly since chain-of-thought
prompting~\citep{Wei2022} and the ReAct
framework~\citep{Yao2023ReAct}.
Toolformer~\citep{SchickToolformer2023} demonstrated that \LLM{}s can
invoke external computational tools via self-supervised learning.
Embodied agents such as PaLM-E~\citep{Driess2023PaLME} and
Voyager~\citep{WangVoyager2024} maintain long-term context about changing
environments.
Self-refinement frameworks —
Self-Refine~\citep{MadaanSelfRefine2023} and
Reflexion~\citep{ShinnReflexion2023} — show that structured feedback
improves \LLM{} outputs iteratively.

\LLM{}s have been deployed as \emph{direct optimizers}:
OPRO~\citep{Yang2024OPRO} treats the \LLM{} as an optimizer over
text prompts;
FunSearch~\citep{Romera2024FunSearch} combines evolutionary search with
\LLM{} code generation for mathematical discovery;
ReEvo~\citep{YeReEvo2024} uses \LLM{} reflections to evolve heuristics
for combinatorial optimization.
The broader intersection of \LLM{}s and evolutionary computation is
surveyed by \citet{WuEvoLLM2024}.

These methods treat the \LLM{} as a \emph{batch optimizer} proposing new
solutions in context.
\AutoSIMP{} uses the \LLM{} in two distinct roles: as a
\emph{configurator} (single-shot parsing of problem description into
structured specification) and as an optional \emph{online controller}
(per-iteration parameter decisions during the solve).
The configurator role is closer to the code-generation paradigm of
\citet{Chen2021Codex} and the scientific agent of \citet{Boiko2023},
while the controller role instantiates the Dynamic Algorithm
Configuration (\textsc{dac}) paradigm of
\citet{BiedenkappDAC2020,AdriaensenAutoDAC2022} without requiring RL
training — the \LLM{} with domain knowledge encoded in its system prompt
acts directly as the policy~\citep{Yang2026LLMController}.

\FloatBarrier
\section{Method}
\label{sec:method}

\begin{figure}[!htbp]
  \centering
  \includegraphics[width=\textwidth]{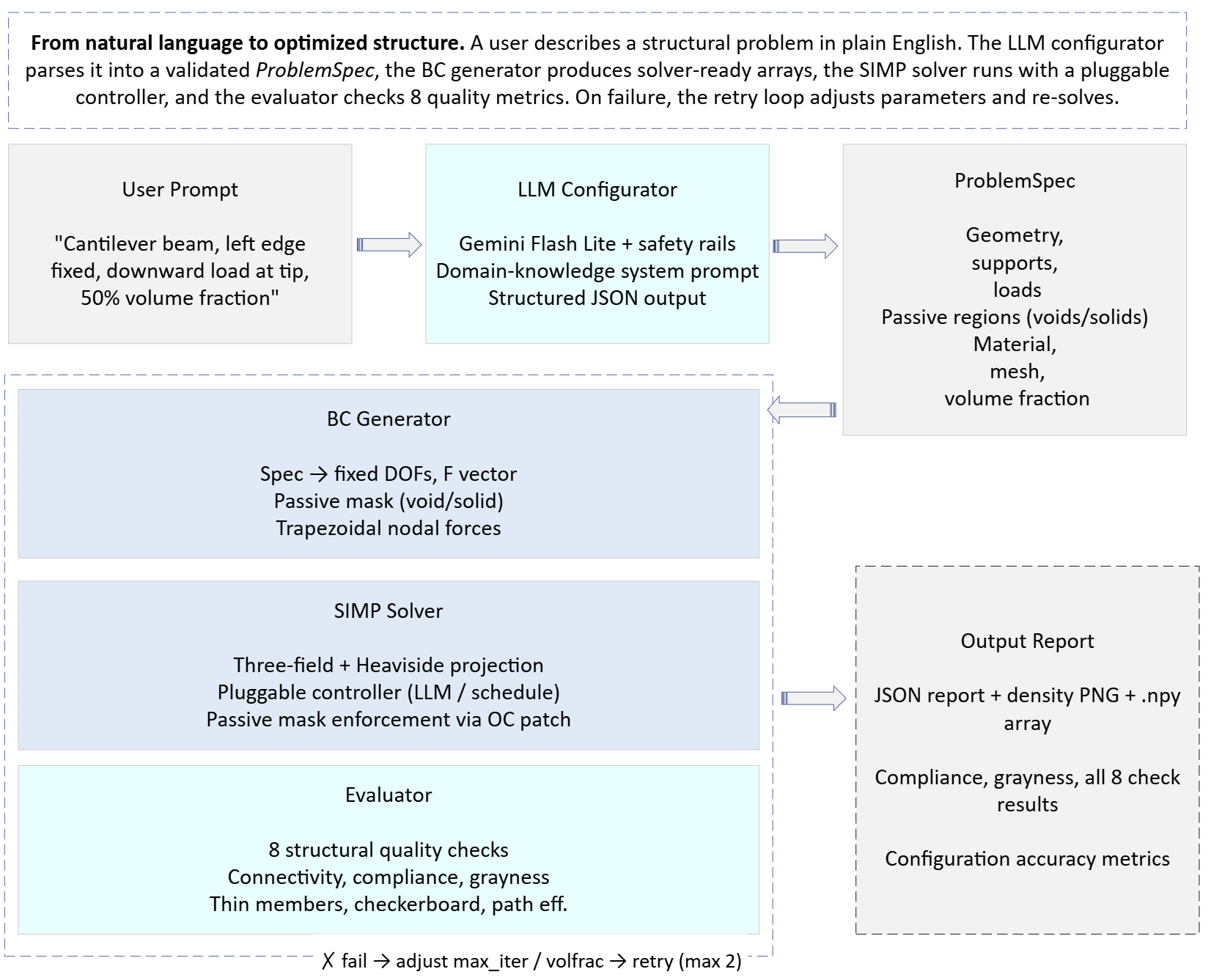}
  \caption{%
    \AutoSIMP{} end-to-end autonomous pipeline.
    A natural-language prompt enters the \LLM{} configurator (Module~1),
    which produces a validated \PSpec{}.
    The BC generator (Module~2) converts this into solver-ready arrays.
    The \SIMP{} solver (Module~3) runs with a pluggable controller.
    The evaluator (Module~4) checks eight quality metrics.
    On failure, the retry loop (Module~5) adjusts parameters and
    re-solves (max 2 retries).
    The solver is consistent with~\citet{Yang2026LLMController}.%
  }
  \label{fig:pipeline}
\end{figure}

\AutoSIMP{} comprises five modules that execute sequentially
(Fig.~\ref{fig:pipeline}): the \LLM{} configurator (\cref{sec:m_config}),
the boundary-condition generator (\cref{sec:m_bc}), the \SIMP{} solver
with pluggable controller (\cref{sec:m_solver}), the structural evaluator
(\cref{sec:m_eval}), and the retry loop (\cref{sec:m_retry}).
Each module is implemented as a standalone Python module; the solver
and LLM controller are consistent with
\citet{Yang2026LLMController}.
Algorithm~\ref{alg:pipeline} gives the top-level orchestration.

\begin{algorithm}[htbp]
\caption{\AutoSIMP{}: End-to-End Autonomous Topology Optimization}
\label{alg:pipeline}
\begin{algorithmic}[1]
\REQUIRE Natural-language prompt $\mathcal{P}$; controller type $\theta$; max retries $R$
\ENSURE Report: density field $\rhotil^*$, compliance $\Comp^*$, evaluation result
\STATE $\mathcal{S} \leftarrow \textsc{Configure}(\mathcal{P})$
  \hfill $\triangleright$ Module 1: \LLM{} configurator
\STATE $(\mathbf{d}_{\text{fix}},\; \mathbf{F},\; \mathbf{m}_{\text{pass}})
  \leftarrow \textsc{GenerateBC}(\mathcal{S})$
  \hfill $\triangleright$ Module 2: BC generator
\STATE $\Comp^* \leftarrow \infty$
\FOR{$r = 1, \dots, R+1$}
  \STATE $(\rhotil, \Comp, \mathcal{H})
    \leftarrow \textsc{SolveSIMP}(\mathcal{S},\;
    \mathbf{d}_{\text{fix}},\; \mathbf{F},\; \mathbf{m}_{\text{pass}},\;
    \theta)$
    \hfill $\triangleright$ Module 3
  \STATE $(\textit{pass},\; \mathbf{c},\; \mathbf{h})
    \leftarrow \textsc{Evaluate}(\rhotil, \Comp, \mathcal{S})$
    \hfill $\triangleright$ Module 4
  \IF{$\Comp < \Comp^*$}
    \STATE $\Comp^* \leftarrow \Comp$;\;
           $\rhotil^* \leftarrow \rhotil$
  \ENDIF
  \IF{\textit{pass} $=$ \TRUE}
    \RETURN $(\rhotil^*,\; \Comp^*,\; \mathbf{c})$
  \ENDIF
  \IF{$r \leq R$}
    \STATE Apply rerun hint $\mathbf{h}$ to $\mathcal{S}$
      \hfill $\triangleright$ Module 5: retry
  \ENDIF
\ENDFOR
\RETURN $(\rhotil^*,\; \Comp^*,\; \mathbf{c})$
  \hfill $\triangleright$ best across attempts
\end{algorithmic}
\end{algorithm}

\subsection{Module 1: LLM Configurator}
\label{sec:m_config}

The configurator transforms a natural-language prompt~$\mathcal{P}$ into
a validated \PSpec{}~$\mathcal{S}$, defined as the tuple
\begin{equation}
  \mathcal{S} \;=\;
  \bigl(L_x, L_y, [L_z],\;
  n_x, n_y, [n_z],\;
  \vf,\;
  \mathcal{B},\;
  \mathcal{L},\;
  \mathcal{R}\bigr),
  \label{eq:spec}
\end{equation}
where $(L_x, L_y, L_z)$ are domain dimensions,
$(n_x, n_y, n_z)$ are element counts,
$\vf$ is the target volume fraction,
$\mathcal{B}$ is a set of support constraints,
$\mathcal{L}$ is a set of load specifications, and
$\mathcal{R}$ is a set of passive-region definitions.
Brackets denote optional 3-D fields.

\paragraph{Support constraints.}
Each $b \in \mathcal{B}$ is either an \emph{edge support}
$(e, \tau)$ specifying an edge name
$e \in \{\text{left}, \text{right}, \text{top}, \text{bottom}\}$
and constraint type
$\tau \in \{\text{fixed}, \text{pin}_x, \text{pin}_y,
\text{roller}_x, \text{roller}_y\}$,
or a \emph{point support} $(\mathbf{x}, \tau)$ at coordinates
$\mathbf{x} = (x, y, [z])$.

\paragraph{Load specifications.}
Each $\ell \in \mathcal{L}$ is either a \emph{point load}
$(\mathbf{x}, \mathbf{f})$ with force vector
$\mathbf{f} = (f_x, f_y, [f_z])$,
or a \emph{distributed load} $(e, q)$ specifying a uniform pressure~$q$
along edge~$e$.

\paragraph{Passive regions.}
Each $r \in \mathcal{R}$ specifies a \emph{circular} region
(centre $(c_x, c_y)$, radius~$r_c$, type $\in \{\text{void}, \text{solid}\}$)
or a \emph{rectangular} region (bounding box, type).

\begin{figure}[!htbp]
  \centering
  \includegraphics[width=\textwidth]{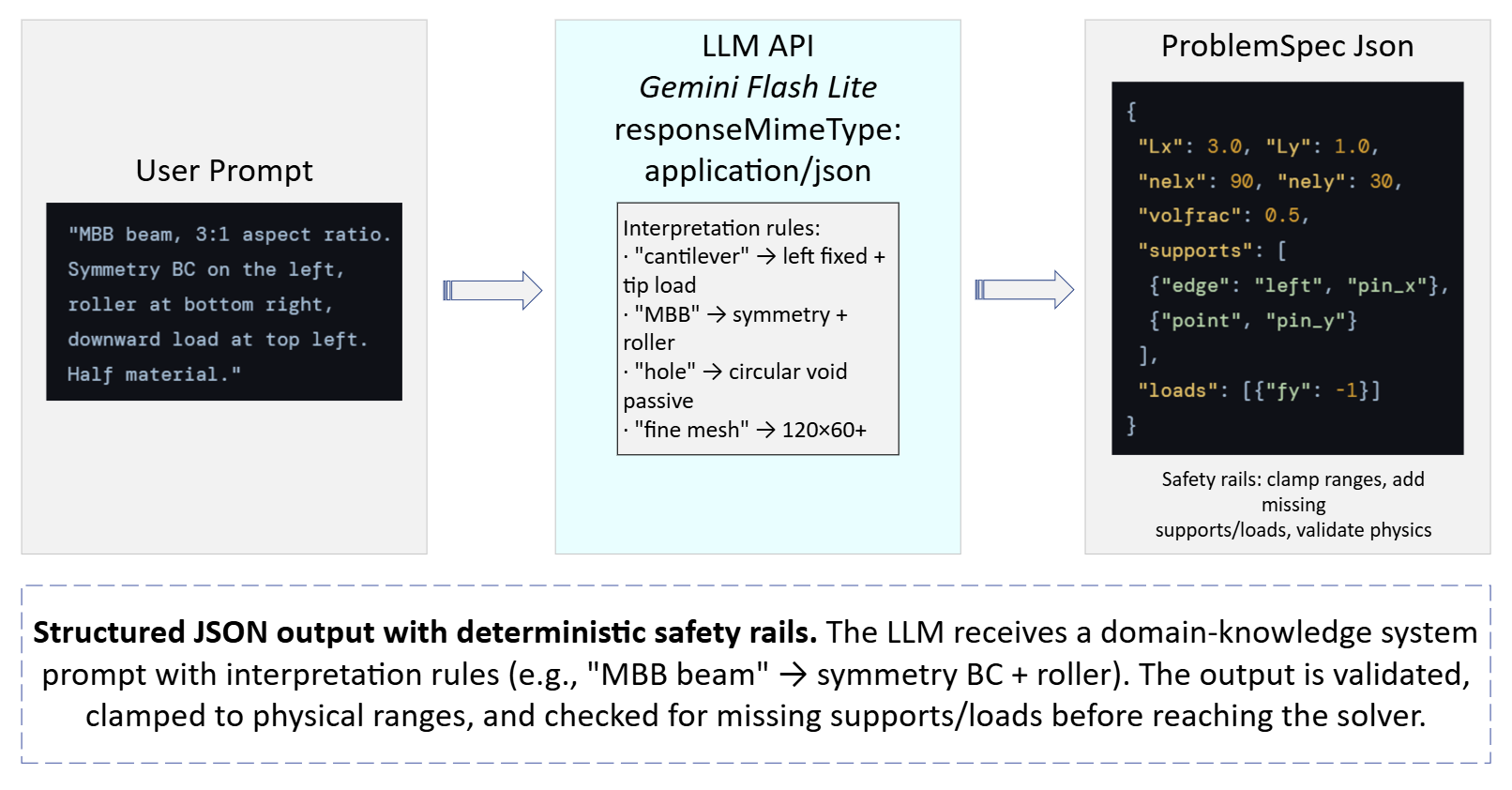}
  \caption{%
    LLM Configurator detail (Module~1).
    A natural-language prompt (left) is processed by Gemini Flash Lite
    with structured JSON output mode and domain-knowledge interpretation
    rules (centre).
    The output is a validated \PSpec{} JSON (right) specifying geometry,
    mesh, supports, loads, and volume fraction.
    Deterministic safety rails clamp ranges, detect missing constraints,
    and validate physical consistency before the specification reaches
    the solver.%
  }
  \label{fig:configurator}
\end{figure}

\paragraph{LLM interface.}
The configurator calls Gemini Flash Lite
(\texttt{gemini-3.1-flash-lite-preview}) at temperature~$0$ with
structured JSON output mode and a response schema that enforces the
\PSpec{} structure.
The system prompt encodes domain-knowledge interpretation rules
(Algorithm~\ref{alg:config} and Fig.~\ref{fig:configurator}):

\begin{algorithm}[htbp]
\caption{LLM Configurator: $\textsc{Configure}(\mathcal{P}) \to \mathcal{S}$}
\label{alg:config}
\begin{algorithmic}[1]
\REQUIRE Prompt $\mathcal{P}$
\ENSURE Validated specification $\mathcal{S}$
\STATE Construct system prompt with domain rules:
\STATE \hspace{1em} ``cantilever'' $\to$ left edge fixed $+$ tip load
\STATE \hspace{1em} ``MBB beam'' $\to$ symmetry BC $+$ roller
\STATE \hspace{1em} ``hole/void'' $\to$ circular passive region
\STATE \hspace{1em} default $\vf = 0.5$, mesh $60 \times 30$
\STATE $\mathcal{S}_{\text{raw}} \leftarrow \textsc{LLM}(\mathcal{P},\;
  \text{schema},\; T{=}0)$
  \hfill $\triangleright$ structured JSON output
\STATE \textbf{Safety rails:}
\STATE \hspace{1em} Clamp $\vf \in [0.1, 0.9]$,\;
  $L_x, L_y > 0$
\STATE \hspace{1em} Reject if $|\mathcal{B}| = 0$
  \hfill $\triangleright$ no supports
\STATE \hspace{1em} Reject if $|\mathcal{L}| = 0$
  \hfill $\triangleright$ no loads
\STATE \hspace{1em} Flag loads applied at fixed DOFs
\STATE \hspace{1em} Adjust mesh if element aspect ratio $> 3$
\IF{LLM call failed}
  \STATE $\mathcal{S} \leftarrow \textsc{RegexFallback}(\mathcal{P})$
    \hfill $\triangleright$ keyword parsing
\ELSE
  \STATE $\mathcal{S} \leftarrow \textsc{Validate}(\mathcal{S}_{\text{raw}})$
\ENDIF
\RETURN $\mathcal{S}$
\end{algorithmic}
\end{algorithm}

\subsection{Module 2: Boundary-Condition Generator}
\label{sec:m_bc}

The BC generator converts a validated $\mathcal{S}$ into three
solver-ready arrays: the set of fixed DOF indices~$\mathbf{d}_{\text{fix}}$,
the global force vector~$\mathbf{F}$, and the passive-element
mask~$\mathbf{m}_{\text{pass}}$.

\paragraph{Fixed DOFs.}
For each edge support $(e, \tau) \in \mathcal{B}$, the generator
identifies all nodes on edge~$e$ and constrains the appropriate
displacement components.
For each point support $(\mathbf{x}, \tau)$, the nearest mesh node is
found:
\begin{equation}
  i^* = \operatorname*{argmin}_{i \in \{1,\dots,N_{\text{node}}\}}
  \| \mathbf{x}_i - \mathbf{x} \|_2.
  \label{eq:snap}
\end{equation}
The constraint mapping is:
$\text{fixed} \to \{u_x, u_y\}$;
$\text{pin}_x \to \{u_x\}$;
$\text{pin}_y \to \{u_y\}$;
$\text{roller}_x \equiv \text{pin}_x$;
$\text{roller}_y \equiv \text{pin}_y$.
For 3-D problems, the expansion includes $u_z$ with analogous logic.

\paragraph{Force vector.}
Point loads are assembled by adding force components to the DOFs of the
nearest node (Eq.~\ref{eq:snap}).
Distributed loads along edge~$e$ of length~$L$ discretized into $n_e$
segments use trapezoidal-rule tributary weighting:
\begin{equation}
  F_i = \begin{cases}
    q \cdot h / 2 & \text{if node } i \text{ is a corner node,} \\
    q \cdot h      & \text{otherwise,}
  \end{cases}
  \qquad h = L / n_e,
  \label{eq:trap}
\end{equation}
ensuring $\sum_{i} F_i = q \cdot L$ exactly.
This was verified DOF-for-DOF against the solver's built-in cantilever
and MBB beam boundary-condition functions.

\paragraph{Passive mask.}
For each passive region $r \in \mathcal{R}$, elements are marked based
on centroid containment:
\begin{equation}
  m_e = \begin{cases}
    \text{void}  & \text{if } \|\mathbf{c}_e - (c_x, c_y)\| < r_c
      \text{ (circular void)}, \\
    \text{solid} & \text{if } \|\mathbf{c}_e - (c_x, c_y)\| < r_c
      \text{ (circular solid)}, \\
    \text{free}  & \text{otherwise},
  \end{cases}
  \label{eq:passive}
\end{equation}
where $\mathbf{c}_e$ is the centroid of element~$e$.
Void elements are frozen at $\rho_{\min} = 10^{-3}$; solid elements
at $\rho = 1.0$.
Enforcement is implemented by patching the solver's \OC{} bisection at
runtime: after each density update, sensitivities on passive elements
are zeroed and densities are reset to their frozen values.
The patch is installed before the solve and removed after completion
via exception-safe cleanup.

\subsection{Module 3: SIMP Solver with Pluggable Controller}
\label{sec:m_solver}

The solver is the three-field \SIMP{} implementation from
\citet{Yang2026LLMController}.
The optimization problem is:
\begin{equation}
\begin{aligned}
  \min_{\boldsymbol{\rho}} \quad & \Comp(\rhotil)
    = \mathbf{F}^T \mathbf{U}(\rhotil) \\
  \text{s.t.} \quad
    & \mathbf{K}(\rhotil)\,\mathbf{U} = \mathbf{F}, \\
    & \frac{1}{|\Omega|}\sum_e \rhotil_e \, v_e \leq \vf, \\
    & \rhotil_e = H_\betaH(\bar{\rho}_e), \quad
      \bar{\rho}_e = \frac{\sum_j w_{ej} \rho_j}{\sum_j w_{ej}},
      \quad \rho_e \in [\rho_{\min}, 1],
\end{aligned}
\label{eq:to}
\end{equation}
where $H_\betaH$ is the Heaviside projection~\citep{Wang2011},
$w_{ej}$ are density filter weights within
radius $\rmin$, and the element stiffness follows the \SIMP{}
interpolation~\citep{Bendsoe1989}.

\paragraph{Solver iteration.}
At each iteration $k$, the solver:
(1)~projects $\boldsymbol{\rho} \to \bar{\boldsymbol{\rho}} \to
\rhotil$ via density filter and Heaviside;
(2)~assembles $\mathbf{K}(\rhotil)$;
(3)~solves $\mathbf{K}\mathbf{U} = \mathbf{F}$ (sparse direct for 2-D;
\AMG{}-preconditioned CG via pyamg~\citep{Bell2023PyAMG} for 3-D);
(4)~computes $\Comp$ and $\partial\Comp/\partial\rho_e$;
(5)~updates $\boldsymbol{\rho}$ via \OC{} with move limit $\dmove$; and
(6)~calls the controller callback.

\paragraph{Controller interface.}
The controller is any Python object that exposes three methods:
an initialization method returning the starting values of
$(\penal, \betaH, \rmin, \dmove)$,
a per-iteration callback that returns updated parameters (or signals
no change), and a finalization method that returns the tail-phase
parameters.
Six controllers are provided (\cref{sec:experiments:controllers}).
All non-fixed controllers share an identical sharpening tail
($\penal = 4.5$, $\betaH = 32$, $\rmin = 1.20$, $\dmove = 0.05$,
$N_{\text{tail}} = 40$).

\paragraph{Validity gate.}
The solver tracks the best snapshot $\rhotil^*_{\text{valid}}$ where
$\penal \geq \penal_{\text{gate}} = 3.0$ and the volume constraint is
satisfied.
The tail restarts from this snapshot.
If no valid snapshot exists (as in the tail-only ablation), the solver
resets to uniform density $\boldsymbol{\rho} = \vf \cdot \mathbf{1}$:
\begin{equation}
  \boldsymbol{\rho}_{\text{tail}} = \begin{cases}
    \rhotil^*_{\text{valid}} & \text{if valid snapshot exists}, \\
    \vf \cdot \mathbf{1}     & \text{otherwise}.
  \end{cases}
  \label{eq:gate}
\end{equation}
This prevents a free warm-start that would confound ablation experiments
(\cref{sec:discussion}).

\subsection{Module 4: Structural Evaluator}
\label{sec:m_eval}

\begin{figure}[!htbp]
  \centering
  \includegraphics[width=\textwidth]{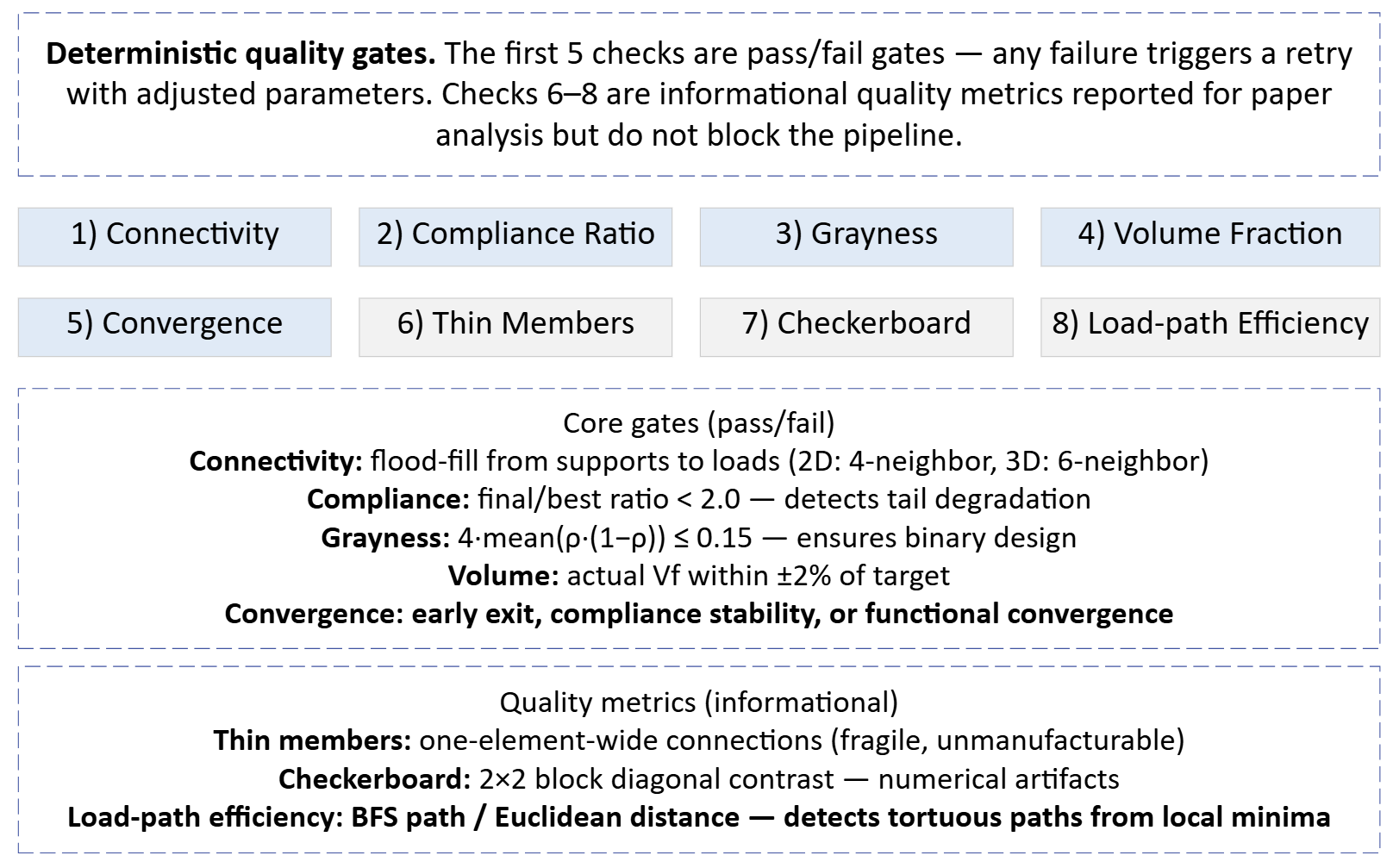}
  \caption{%
    Structural quality evaluator (Module~4).
    The first five checks are pass/fail gates — any failure triggers a
    retry with adjusted parameters.
    Checks 6--8 are informational quality metrics reported for analysis
    but do not block the pipeline.
    Core gates (blue) ensure structural validity; informational metrics
    (grey) characterise design quality.%
  }
  \label{fig:evaluator}
\end{figure}

The evaluator performs eight deterministic checks on the solver output
(Table~\ref{tab:checks} and Fig.~\ref{fig:evaluator}).
Five are pass/fail gates; three are informational metrics.
A design \emph{passes} if and only if all five gates are satisfied.

\begin{table}[!htbp]
\centering
\caption{Eight structural quality checks. Gates 1--5 determine pass/fail;
metrics 6--8 are informational.}
\label{tab:checks}
\small
\begin{tabular}{clll}
\toprule
\# & Check & Criterion & Type \\
\midrule
1 & Connectivity & Flood fill from supports to loads $\geq 0.99$ & Gate \\
2 & Compliance ratio & $\Comp_{\text{final}} / \Comp_{\text{best}} < 2.0$ & Gate \\
3 & Grayness & $\Gray = 4\,\overline{\rho(1-\rho)} \leq 0.15$ & Gate \\
4 & Volume fraction & $|V_{f,\text{actual}} - \vf| \leq 0.02$ & Gate \\
5 & Convergence & Stability or early exit & Gate \\
6 & Thin members & Fraction of 1-element-wide connections & Metric \\
7 & Checkerboard & $2 \times 2$ block diagonal contrast & Metric \\
8 & Load-path eff. & BFS path / Euclidean distance & Metric \\
\bottomrule
\end{tabular}
\end{table}

\paragraph{Connectivity (Gate~1).}
A flood fill from support nodes through solid elements
($\rhotil_e > 0.5$) using 4-connectivity (2-D) or 6-connectivity (3-D).
The connectivity fraction is
\begin{equation}
  f_{\text{conn}} =
  \frac{|\{e : \rhotil_e > 0.5 \text{ and } e \text{ reached}\}|}
       {|\{e : \rhotil_e > 0.5\}|}.
  \label{eq:conn}
\end{equation}

\paragraph{Grayness (Gate~3).}
The grayness index is defined as
\begin{equation}
  \Gray(\rhotil) = \frac{4}{N_e} \sum_{e=1}^{N_e}
    \rhotil_e \,(1 - \rhotil_e),
  \label{eq:gray}
\end{equation}
where $\Gray = 0$ for a fully binary design and $\Gray = 1$ for
uniform $\rhotil = 0.5$.

\paragraph{Rerun hints.}
When a gate fails, the evaluator produces a structured hint~$\mathbf{h}$:
\begin{equation}
  \mathbf{h} = \begin{cases}
    \{N \leftarrow 1.3 \times N\} &
      \text{convergence failure}, \\
    \{\vf \leftarrow \vf + \Delta\vf\} &
      \text{volume violation by } \Delta\vf, \\
    \{N \leftarrow 1.3 \times N\} &
      \text{grayness failure}.
  \end{cases}
  \label{eq:hint}
\end{equation}

\subsection{Module 5: Retry Loop}
\label{sec:m_retry}

The orchestrator wraps Modules~2--4 in a retry loop
(Algorithm~\ref{alg:pipeline}, lines~4--16) with a configurable
maximum retry count~$R$ (default~2).
On each failed attempt, the evaluator's hint~$\mathbf{h}$
(Eq.~\ref{eq:hint}) modifies the \PSpec{} or solver parameters before
re-entering the solve.
If no specific hint is available, the iteration budget is increased by
$30\%$ as a default recovery strategy.
The loop tracks the best compliance across all attempts and returns the
overall best result.

In practice, with the deterministic schedule controller, all~10
pipeline test cases pass on the first attempt
(Table~\ref{tab:retry}), so the retry mechanism serves as a safety net
for edge cases rather than a routinely exercised component.

\subsection{Module 6: Interactive Web Interface}
\label{sec:m_demo}

\begin{figure}[!htbp]
  \centering
  \begin{subfigure}[t]{0.45\textwidth}
    \centering
    \includegraphics[width=0.8\textwidth]{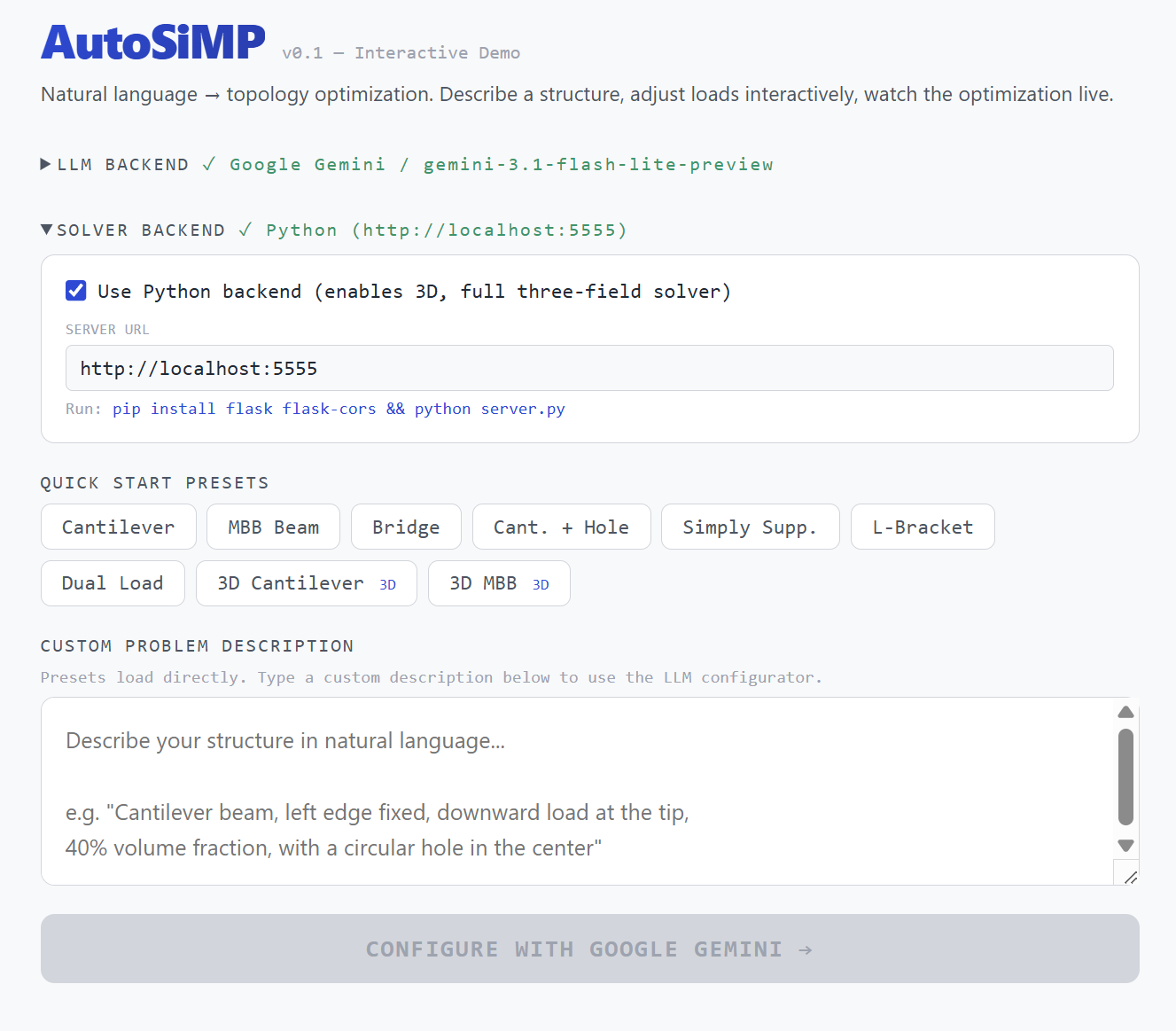}
    \caption{Input page: multi-provider \LLM{} configuration, Python
    backend toggle, nine presets (seven 2-D, two 3-D), and
    natural-language prompt field.}
    \label{fig:demo:input}
  \end{subfigure}\hfill
  \begin{subfigure}[t]{0.45\textwidth}
    \centering
    \includegraphics[width=0.8\textwidth]{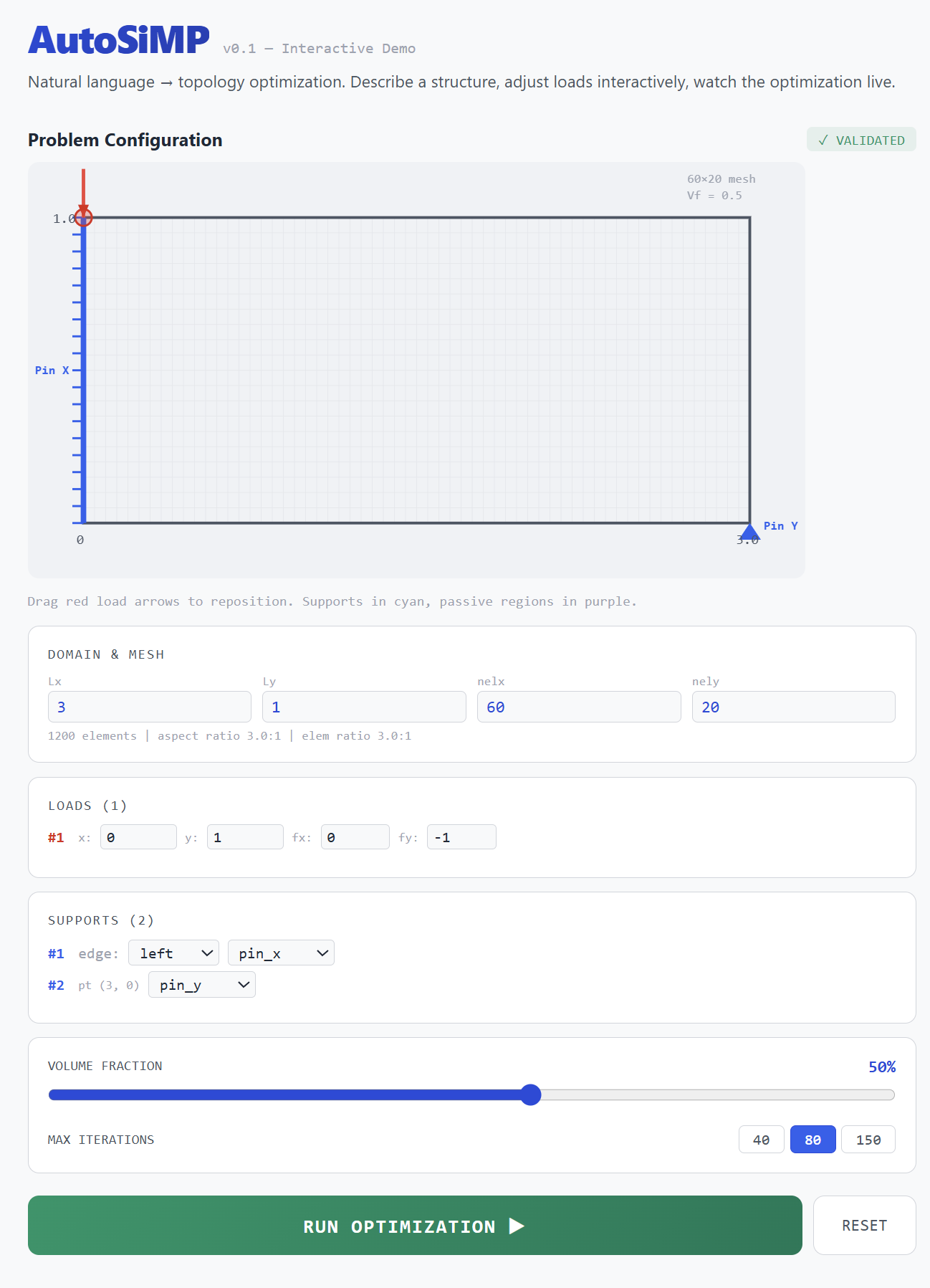}
    \caption{Interactive preview: draggable load arrows, editable domain/mesh
    parameters, support-type dropdowns, volume-fraction slider, and
    iteration budget selection.}
    \label{fig:demo:preview}
  \end{subfigure}\\[8pt]
  \begin{subfigure}[t]{0.45\textwidth}
    \centering
    \includegraphics[width=0.8\textwidth]{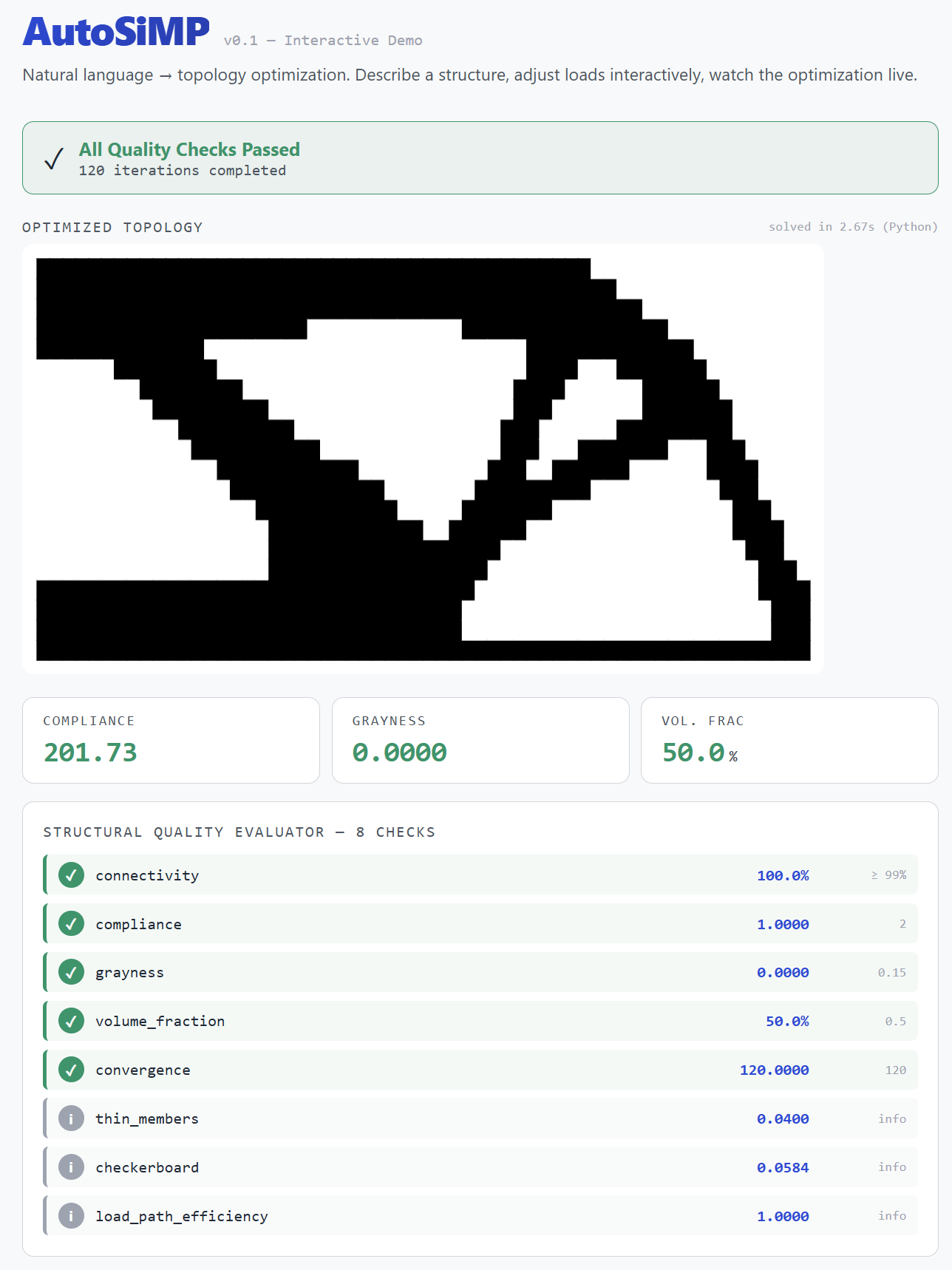}
    \caption{Optimized \MBB{} beam (60$\times$20, $\vf=0.5$, 120
    iterations, Python backend): binary topology with all eight
    quality checks passed, solved in 2.67\,s.}
    \label{fig:demo:result}
  \end{subfigure}\hfill
  \begin{subfigure}[t]{0.45\textwidth}
    \centering
    \includegraphics[width=0.8\textwidth]{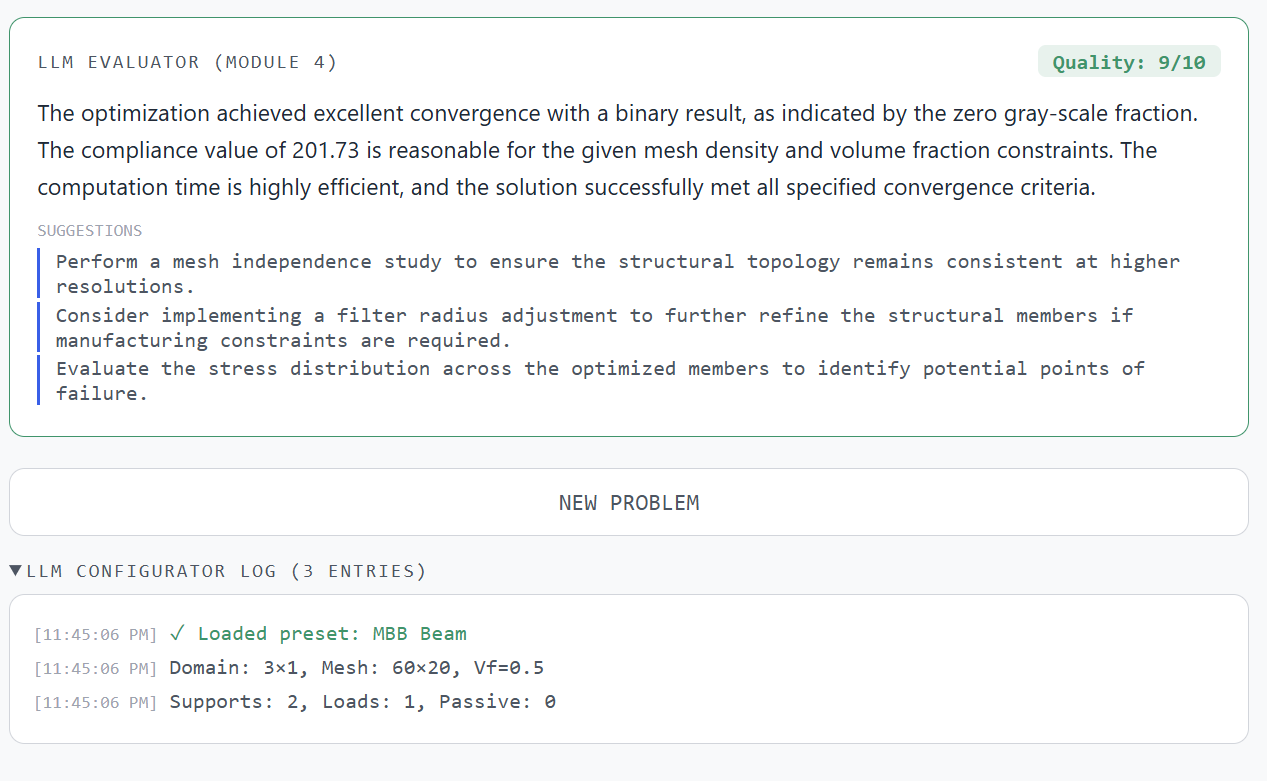}
    \caption{\LLM{} Evaluator (Module~4): qualitative assessment
    with quality score (9/10) and actionable suggestions, generated
    by Gemini~3.1 Flash Lite.}
    \label{fig:demo:eval}
  \end{subfigure}
  \caption{Interactive web demo for \AutoSIMP{}.
  The interface implements the complete five-module pipeline in a
  browser, supporting both client-side JavaScript solving (instant 2-D)
  and Python backend solving (production-quality 2-D/3-D with
  three-field \SIMP{} and 40-iteration sharpening tail).
  No installation is required beyond a web browser; the Python backend
  is optional and consists of a single Flask file.
  None of the systems in Table~\ref{tab:related} provides an
  interactive visual interface for topology optimization from natural
  language.}
  \label{fig:demo}
\end{figure}

To demonstrate the complete autonomous pipeline in a visual, accessible
format, we provide a browser-based interactive demo
(Fig.~\ref{fig:demo}).
The interface implements a five-stage workflow:
(1)~natural-language input with configurable \LLM{} backend
(Google~Gemini, OpenAI, Anthropic~Claude, or any OpenAI-compatible
endpoint);
(2)~\LLM{} configurator with visible safety-rail log showing each
parsing step, validation check, and fallback decision;
(3)~interactive problem preview with draggable load points and fully
editable specification fields (domain dimensions, mesh resolution,
support types, load positions and magnitudes, volume fraction, and
iteration budget);
(4)~live optimization with real-time density visualization and
continuation-parameter display ($\penal$, $\betaH$, compliance,
change); and
(5)~results with the eight-check evaluator panel, optional \LLM{}
qualitative assessment with a quality score, and compliance convergence
history.

The demo operates in two solver modes.
In \emph{browser mode}, a JavaScript implementation of the three-field
\SIMP{} solver with Heaviside projection and the paper's four-phase
continuation schedule (\cref{sec:experiments:controllers}) runs entirely
client-side, enabling instant 2-D optimization without server
infrastructure.
In \emph{backend mode}, a lightweight Flask API
($\sim$170 lines) wraps the production Python solver unchanged,
supporting both 2-D and 3-D problems.
Three-dimensional results are rendered as interactive isosurfaces via
marching cubes in Three.js with mouse-driven rotation and zoom, plus
orthogonal X-ray density projections (mean along each axis).
Nine presets are included (seven 2-D, two 3-D), covering the
benchmark suite from \cref{sec:experiments:benchmarks}.
All \LLM{} and solver settings are persisted in browser local storage.

The interface is implemented as a single React component
($\sim$2{,}300~lines) served by Vite, with Three.js loaded from CDN
for 3-D visualization.
No competitor system in Table~\ref{tab:related} provides an
interactive visual interface for topology optimization from natural
language.

\FloatBarrier
\section{Experimental Setup}
\label{sec:experiments}

We evaluate \AutoSIMP{} on three complementary axes, each targeting a
different component of the autonomous pipeline.
\emph{\Cref{tab:pipeline}} measures the \textbf{configurator}: does the \LLM{}-parsed
specification produce a topology comparable to one configured by a domain
expert?
\emph{\Cref{tab:controllers}} measures the \textbf{controller}: across a large benchmark
suite, how do six continuation strategies compare in compliance and
reliability?
\emph{\Cref{tab:retry}} measures the \textbf{end-to-end loop}: when the configurator
and evaluator operate together, does the system achieve 100\% pass rate?
All experiments use a single random seed (42) and are fully deterministic
for the non-\LLM{} controllers.

\subsection{Benchmark Suite}
\label{sec:experiments:benchmarks}

\paragraph{2-D problems (17).}
The benchmark suite comprises 17 two-dimensional problems spanning five
categories:
(i)~standard cantilevers at three mesh resolutions
  (60$\times$30, 120$\times$40) and three volume fractions (0.3, 0.4, 0.6);
(ii)~\MBB{} beams at two resolutions (90$\times$30, 150$\times$50);
(iii)~problems with passive regions — a cantilever with a circular void,
  a cantilever with two circular voids, and a cantilever with a rectangular
  solid insert;
(iv)~non-standard geometries — an L-bracket (60$\times$60 with
  upper-right quadrant voided), a bridge with bottom-edge support and
  distributed top load, a simply supported beam, and a deep cantilever
  (1:2 aspect ratio); and
(v)~a high-aspect-ratio cantilever (6:1, 120$\times$20).
All problems use compliance minimization with a volume constraint.
Boundary conditions are specified via the \PSpec{} dataclass and converted
to solver-ready arrays by the BC generator; the solver itself receives
only DOF indices, force vectors, and passive masks.

\paragraph{3-D problems (2).}
Two three-dimensional problems are included to validate generalization:
a 3-D cantilever and a 3-D \MBB{} beam, both on a 30$\times$15$\times$8
mesh (3{,}600 elements, $\approx$13{,}000 DOFs) with $\vf = 0.4$.
The solver uses \AMG{}-preconditioned conjugate gradient via
pyamg~\citep{Bell2023PyAMG} for the 3-D linear systems.

\paragraph{Pipeline test cases (10).}
For the configurator evaluation (\cref{tab:pipeline}), 10 problems are selected from
the benchmark suite and paired with natural-language prompts.
Each prompt describes the problem in plain English without solver-specific
terminology — \eg, ``\textit{Cantilever beam, left edge fixed, downward
load at mid-right, 50\% material, circular hole at center for a pipe}''
rather than ``fixed DOFs on left edge, $F_y = -1$ at node $(nelx, nely/2)$,
void mask for circle at $(Lx/2, Ly/2, r=0.2)$''.
Ground-truth \PSpec{} objects are provided for each case, enabling
field-by-field comparison of the \LLM{} output.

\subsection{Controllers}
\label{sec:experiments:controllers}

Six continuation controllers are compared, all sharing an identical
40-iteration sharpening tail (STANDARD\_TAIL: $\penal = 4.5$,
$\betaH = 32$, $\rmin = 1.20$, $\dmove = 0.05$, restart from best
valid snapshot).
This design ensures that compliance differences are attributable solely to
the exploration phase.

\begin{enumerate}
  \item \textbf{LLM agent.}
    Gemini Flash Lite (\texttt{gemini-3.1-flash-lite-preview}) with
    \DNC{} interface.
    Observes solver state every $k$ iterations and outputs
    $(\penal, \betaH, \rmin, \dmove, \text{restart})$ with a grayness
    gate~\citep{Yang2026LLMController}.

  \item \textbf{Schedule.}
    Identical four-stage phase structure as the \LLM{} agent (exploration
    $\to$ penalization $\to$ sharpening $\to$ convergence) but with
    deterministic, pre-computed parameter values.
    No \LLM{} API calls.
    This is the key ablation: it isolates the contribution of the phase
    structure from the \LLM{}'s adaptive decisions.

  \item \textbf{Expert heuristic.}
    A manually designed step-wise $\penal$ ramp with $\betaH$ raised only
    after $\penal \geq 3.0$, and a safe restart on valid best + severe
    compliance spike.

  \item \textbf{Three-field continuation.}
    Linear $\penal$ ramp ($1 \to 4.5$ over 30 iterations), geometric
    $\betaH$ doubling every 10 iterations, late $\rmin$ tightening.
    Represents the standard academic continuation
    baseline~\citep{Wang2011,Lazarov2016}.

  \item \textbf{Tail-only.}
    Zero exploration: forces $\penal = 1.0$ throughout the main loop,
    keeping the validity gate unsatisfied ($\penal < 3.0$).
    The tail then starts from \emph{uniform density} (not from a
    warmed-up snapshot), so it must produce a topology from scratch
    in 40 iterations.
    This ablation confirms that exploration genuinely matters.

  \item \textbf{Fixed (no continuation).}
    No parameter changes, no tail.
    Runs 300 iterations at $\penal = 3.0$, $\betaH = 1.0$.
    Produces gray (unconverged) designs; serves as the true
    no-intervention baseline.
\end{enumerate}

\subsection{Evaluation Protocol}
\label{sec:experiments:protocol}

Each solve is evaluated by the eight-check structural evaluator.
The five pass/fail gates are:
(1)~\textbf{connectivity} — flood-fill from supports to loads
  (4-neighbor in 2-D, 6-neighbor in 3-D);
(2)~\textbf{compliance ratio} — final/best $< 2.0$, detecting tail
  degradation;
(3)~\textbf{grayness} — $4 \cdot \text{mean}(\rho(1-\rho)) \leq 0.15$;
(4)~\textbf{volume fraction} — within $\pm 2\%$ of target; and
(5)~\textbf{convergence} — early exit, compliance stability
  (relative range $< 0.005$), or functional convergence.
Three informational metrics are also recorded:
(6)~thin-member fraction,
(7)~checkerboard index, and
(8)~load-path efficiency.
A solve \emph{passes} if and only if all five gates are satisfied.

\subsection{Experimental Axes}
\label{sec:experiments:axes}

\paragraph{Configuration accuracy (\cref{tab:pipeline}).}
For each of the 10 pipeline test cases, the \LLM{} configurator parses
the natural-language prompt into a \PSpec{}.
We measure:
(a)~\emph{field accuracy} — how many of the 6 specification fields
  (geometry, mesh, supports, loads, passive regions, volume fraction) match
  the ground truth; and
(b)~\emph{compliance penalty} — the relative compliance difference
  between the \LLM{}-configured solve and the ground-truth solve, both
  using the same controller (LLM agent, 300 iterations).

\paragraph{Controller comparison (\cref{tab:controllers}).}
All 17 2-D problems and 2 3-D problems are solved with each of the 6
controllers at 300 iterations (seed~42).
We report pass rate, mean compliance, median compliance, mean grayness,
and mean wall-clock time per controller.
The per-problem compliance matrix enables direct comparison of controller
quality across the full benchmark suite.

\paragraph{Retry recovery (\cref{tab:retry}).}
All 10 pipeline test cases are run through the full \AutoSIMP{} orchestrator
with the schedule controller: once with no retries (single-shot) and once
with up to 2 retries.
This measures whether the evaluator's diagnostic and retry mechanism
recovers failures.

\paragraph{Demo validation.}
All nine presets (seven 2-D, two 3-D) were verified through the
interactive web demo (Section~\ref{sec:m_demo}) using the Python
backend with the schedule controller at 80--150 iterations.
The browser-side JavaScript solver produces topologies qualitatively
consistent with the Python solver for standard problems (cantilever,
\MBB{}) but with higher grayness due to the absence of the 40-iteration
sharpening tail.
All 3-D results in the demo use the production Python solver
exclusively.

\FloatBarrier
\section{Results}
\label{sec:results}

\subsection{Configuration Accuracy}
\label{sec:results:table1}

Table~\ref{tab:pipeline} reports the configurator's performance across
10 diverse structural problems.
The \LLM{} configurator produces structurally valid specifications on
all 10 cases (9 with 6/6 matching fields, 1 with 5/6), with a median
compliance penalty of $+0.3\%$ relative to expert-configured ground
truth (excluding one case where the \LLM{} \emph{controller}, not the
configurator, caused a solver failure).

\begin{table}[!htbp]
\centering
\caption{Configuration accuracy: ground-truth (GT) vs.\ \LLM{}-configured
compliance. Both use the \LLM{} controller with 300 iterations. Config
reports the number of matching specification fields (out of 6).}
\label{tab:pipeline}
\small
\begin{tabular}{lrrrl}
\toprule
Problem & GT $\Comp$ & LLM $\Comp$ & Penalty & Config \\
\midrule
cantilever\_basic       &   63.69 &     63.69 &   0.0\% & 6/6~\checkmark \\
mbb\_beam               &  185.16 &    186.24 &  +0.6\% & 6/6~\checkmark \\
bridge                  &   17.95 &      ---  &  FAIL$^\dagger$ & 5/6~\checkmark \\
cantilever\_hole        &   76.24 &     76.24 &   0.0\% & 6/6~\checkmark \\
deep\_beam\_shear       &    5.50 &      5.50 &   0.0\% & 6/6~\checkmark \\
simply\_supported       &   12.88 &     13.86 &  +7.7\% & 6/6~\checkmark \\
cantilever\_low\_vf     &  110.50 &    110.50 &   0.0\% & 6/6~\checkmark \\
dual\_load              &  251.99 &    268.03 &  +6.4\% & 6/6~\checkmark \\
lbracket                &   11.30 &     24.75 & +119\%  & 6/6~\checkmark \\
high\_aspect            & 1151.05 &   1141.11 &  $-$0.9\% & 6/6~\checkmark \\
\midrule
\multicolumn{2}{l}{Median penalty (excl.\ bridge)} & & \textbf{+0.3\%} & \\
\multicolumn{2}{l}{GT pass rate}                    & & 100\% & \\
\multicolumn{2}{l}{LLM pass rate}                   & & 80\%  & \\
\bottomrule
\end{tabular}

\smallskip
\noindent
$^\dagger$Bridge: the configurator correctly parsed 5/6 fields but the
\LLM{} \emph{controller} destabilized the solve (compliance blowup to
$1.57 \times 10^7$). This is a controller failure, not a configuration error.
\end{table}

\begin{figure}[!htbp]
\centering
\begin{tabular}{@{}c@{\hspace{2pt}}c@{\hspace{8pt}}c@{\hspace{2pt}}c@{\hspace{8pt}}c@{}}
 & GT & & LLM & Penalty \\[3pt]
\rotatebox{90}{\small\hspace{4pt}cantilever} &
\includegraphics[width=0.18\textwidth]{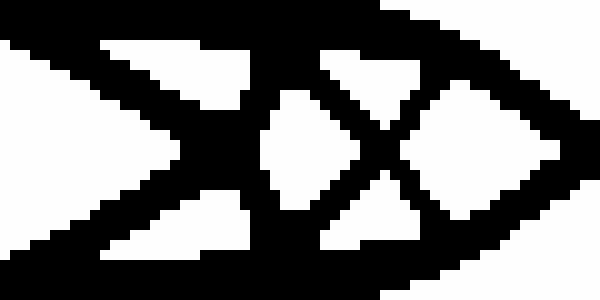} &  &
\includegraphics[width=0.18\textwidth]{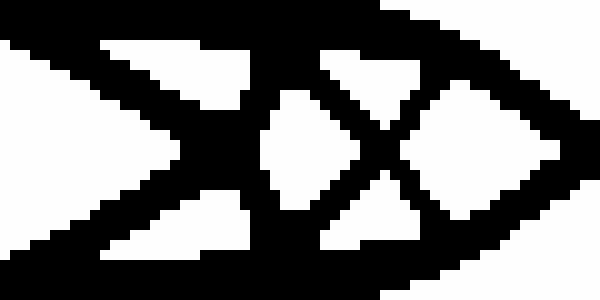} &
\small 0.0\% \\[3pt]
\rotatebox{90}{\small\hspace{4pt}cant.+hole} &
\includegraphics[width=0.18\textwidth]{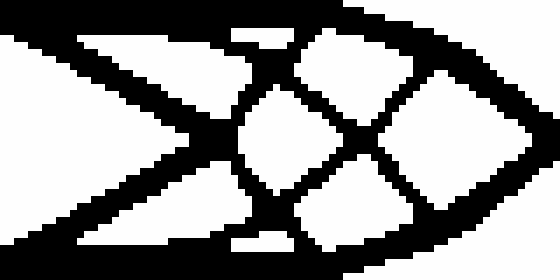} &  &
\includegraphics[width=0.18\textwidth]{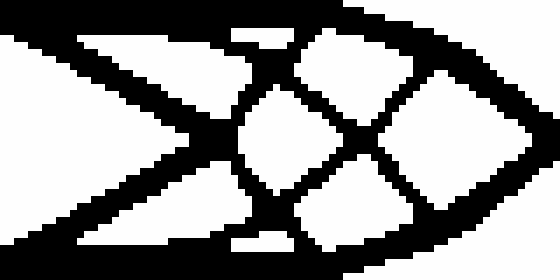} &
\small 0.0\% \\[3pt]
\rotatebox{90}{\small\hspace{8pt}MBB} &
\includegraphics[width=0.18\textwidth]{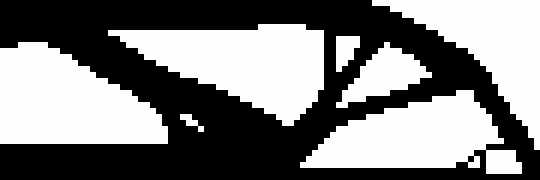} &  &
\includegraphics[width=0.18\textwidth]{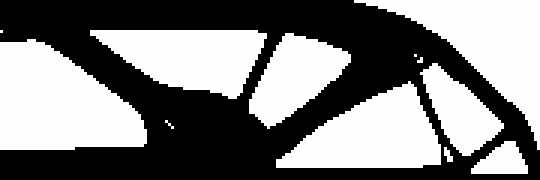} &
\small +0.6\% \\[3pt]
\rotatebox{90}{\small\hspace{0pt}simply supp.} &
\includegraphics[width=0.18\textwidth]{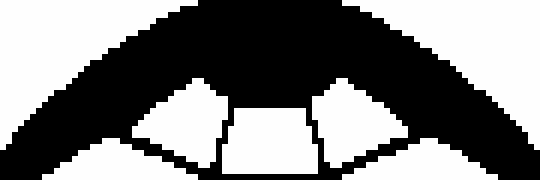} &  &
\includegraphics[width=0.18\textwidth]{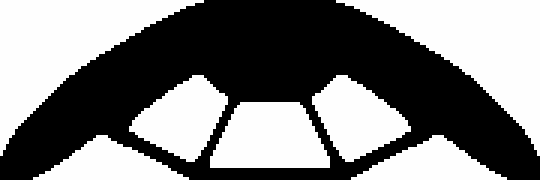} &
\small +7.7\% \\[3pt]
\rotatebox{90}{\small\hspace{4pt}L-bracket} &
\includegraphics[width=0.18\textwidth]{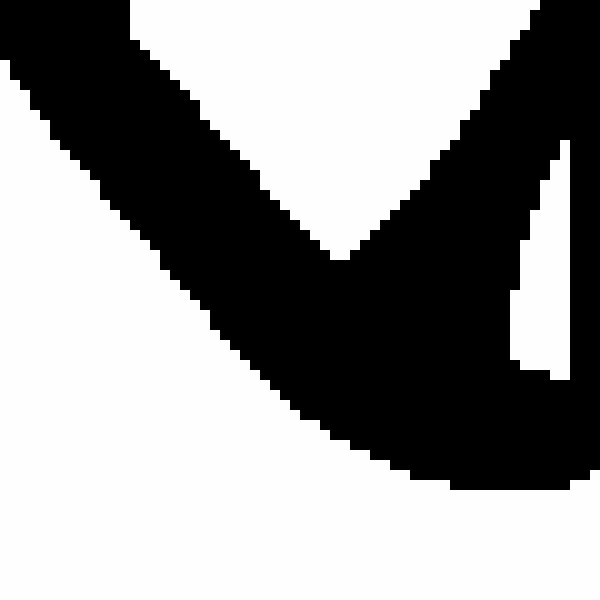} &  &
\includegraphics[width=0.18\textwidth]{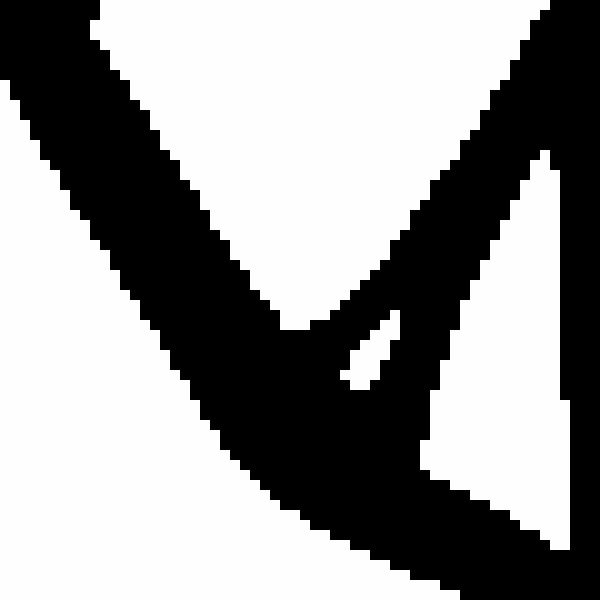} &
\small +119\% \\[2pt]
\end{tabular}
\caption{Configuration accuracy (\cref{tab:pipeline}): ground-truth (GT) vs.\
\LLM{}-configured topologies, both solved with the schedule controller
at 300 iterations. The first three pairs are visually identical
(0.0--0.6\% penalty). The simply supported beam shows minor differences
(+7.7\%). The L-bracket is the one genuine outlier (+119\%): the
configurator placed the load at a different y-coordinate, producing a
structurally different topology.}
\label{fig:table1}
\end{figure}

Five of the nine valid cases achieve 0.0\% penalty — the \LLM{}-configured
and ground-truth specifications produce identical topologies
(Fig.~\ref{fig:table1}).
The simply supported beam ($+7.7\%$) and dual-load cantilever ($+6.4\%$)
show minor penalty from slight differences in how the configurator
interprets load position or support type.
The L-bracket ($+119\%$) is the one genuine outlier: the configurator
produced a correct specification (6/6 fields match the expected format)
but placed the horizontal load at a different y-coordinate than the ground
truth, resulting in a structurally different — and worse — topology
(Fig.~\ref{fig:table1}, bottom row).
The high-aspect cantilever achieves $-0.9\%$ penalty, indicating that the
\LLM{}-configured version found a marginally \emph{better} topology than
the ground truth on this problem.

\subsection{Controller Comparison}
\label{sec:results:table2}

Table~\ref{tab:controllers} summarizes the six-controller comparison
across 17 two-dimensional benchmark problems.

\begin{table}[!htbp]
\centering
\caption{Controller comparison (summary): 17 2-D problems, 300 iterations, seed~42.
All controllers share an identical 40-iteration STANDARD\_TAIL.}
\label{tab:controllers}
\small
\begin{tabular}{lrrrrl}
\toprule
Controller & Pass\% & Mean $\Comp$ & Med.\ $\Comp$ & Mean $\Gray$ & Mean $t$ (s) \\
\midrule
LLM agent       & 76.5\% & 159.60 & \textbf{76.24} & 0.000 & 130.0 \\
Schedule        & \textbf{100\%} & 170.10 & 77.35 & 0.000 & 106.2 \\
Expert          & 88.2\% & 161.00 & 76.44 & 0.000 & 122.1 \\
Three-field     & 88.2\% & 161.28 & 76.75 & 0.000 & 117.3 \\
Tail-only       & 100\%  & 452.62 & 211.78 & 0.002 & 22.9 \\
Fixed           & 35.3\% & 175.73 & 86.74 & 0.169 & 91.0 \\
\bottomrule
\end{tabular}
\end{table}

Table~\ref{tab:perproblem} provides the full per-problem compliance
matrix, enabling direct comparison across all 17 benchmarks.
The best (lowest) compliance per problem is shown in bold.

\begin{table}[!htbp]
\centering
\caption{Per-problem compliance matrix: 17 2-D problems $\times$ 6
controllers, 300 iterations, seed~42.
Best compliance per problem in \textbf{bold}; ties are bolded for all
matching controllers.
All controllers except Fixed share an identical 40-iteration
STANDARD\_TAIL (Fixed has no tail).}
\label{tab:perproblem}
\small
\setlength{\tabcolsep}{5pt}
\begin{tabular}{lrrrrrr}
\toprule
Problem & LLM & Schedule & Expert & Three-field & Tail-only & Fixed \\
\midrule
\multicolumn{7}{l}{\textit{Standard benchmarks}} \\
cantilever\_60$\times$30       &   63.69 &   64.46 & \textbf{63.13} &   63.27 &  141.17 &   68.54 \\
cantilever\_120$\times$40      & \textbf{171.65} &  177.42 &  172.63 &  173.80 &  580.81 &  181.03 \\
mbb\_90$\times$30              & \textbf{185.16} &  203.11 &  190.34 &  191.07 &  337.66 &  204.81 \\
mbb\_150$\times$50             & \textbf{186.05} &  202.12 &  188.12 &  189.22 &  309.73 &  196.80 \\
\midrule
\multicolumn{7}{l}{\textit{Volume-fraction variations}} \\
cantilever\_vf30 (60$\times$30)     &  110.50 &  112.67 & \textbf{107.68} &  108.38 &  602.35 &  128.26 \\
cantilever\_vf40 (60$\times$30)     & \textbf{78.47} &   83.08 &   79.49 &   79.45 &  211.78 &   88.04 \\
cantilever\_vf60 (60$\times$30)     & \textbf{53.93} &   54.72 &   54.16 &   54.07 &   69.52 &   57.08 \\
\midrule
\multicolumn{7}{l}{\textit{Non-standard BCs \& distributed loads}} \\
bridge\_120$\times$30          & \textbf{17.95} &   23.92 &   19.67 &   18.88 &  552.55 &   24.72 \\
cantilever\_dual\_load (80$\times$40) & \textbf{251.99} &  260.11 &  254.22 &  255.46 &  569.30 &  279.25 \\
cantilever\_shear (60$\times$30)   & \textbf{63.60} &   65.37 &   64.02 &   64.51 &  118.08 &   69.44 \\
lbracket\_60$\times$60         & \textbf{11.30} &   11.42 & \textbf{11.30} & \textbf{11.30} &   13.75 &   11.80 \\
simply\_supported (90$\times$30)   &   12.88 &   13.03 & \textbf{12.87} &   12.90 &   13.85 &   13.48 \\
\midrule
\multicolumn{7}{l}{\textit{Passive regions}} \\
cantilever\_hole (80$\times$40)     & \textbf{76.24} &   77.35 &   76.44 &   76.75 &  125.58 &   86.74 \\
cantilever\_two\_holes (90$\times$30) &   210.53 &  217.62 & \textbf{210.49} &  212.71 &  692.06 &  233.85 \\
cantilever\_solid\_insert (80$\times$40) & \textbf{62.75} &   64.22 &   63.07 &   62.99 &  161.78 &   67.00 \\
\midrule
\multicolumn{7}{l}{\textit{Aspect-ratio extremes}} \\
cantilever\_6:1 (120$\times$20)   & \textbf{1151.05} & 1255.60 & 1163.93 & 1161.56 & 3189.01 & 1270.90 \\
deep\_cantilever (30$\times$60)     & \textbf{5.50} &    5.51 &    5.51 & \textbf{5.50} &    5.56 &    5.70 \\
\bottomrule
\end{tabular}
\end{table}

Several findings emerge from this comparison
(Tables~\ref{tab:controllers} and~\ref{tab:perproblem}).

\paragraph{Per-problem patterns.}
The per-problem matrix (Table~\ref{tab:perproblem}) reveals that no single
controller dominates across all 17~problems.
The \LLM{} controller achieves the lowest or tied-lowest compliance on
13 of 17~problems (11 sole wins, 2 ties), primarily on multi-load,
distributed-load, higher-resolution meshes, and passive-region cases.
The expert heuristic wins or ties on 5~problems (standard cantilever,
volume-fraction variants, and simply supported beam), and the three-field
continuation ties on 2~problems (L-bracket, deep cantilever).
The schedule controller never achieves the best compliance on any single
problem, yet its aggregate performance is within $1.5\%$ of the best
median—confirming that its advantage lies in \emph{consistency} rather
than peak performance.
The tail-only controller is $2\times$--$30\times$ worse across all
problems, with the largest gaps on higher-resolution meshes
(cantilever\_120$\times$40: $580.81$ vs.\ $171.65$) and the bridge
($552.55$ vs.\ $17.95$), where topological complexity demands
extended exploration.

\paragraph{The quality--reliability trade-off.}
The \LLM{} controller achieves the lowest median compliance (76.24) but the
lowest pass rate among active controllers (76.5\%), due to API-induced
parameter oscillations on larger meshes.
The schedule controller achieves 100\% pass rate with only $+1.5\%$ higher
median compliance (77.35).
This trade-off is the central practical finding: the deterministic schedule
is the recommended default for production use, while the \LLM{} controller
is valuable for exploratory runs where the user can tolerate occasional
failures in exchange for marginally better topologies.

\begin{figure}[!htbp]
\centering
\setlength{\tabcolsep}{1pt}
\begin{tabular}{@{}cccccc@{}}
 & \small LLM & \small Schedule & \small Expert & \small Tail-only & \small Fixed \\[2pt]
\rotatebox{90}{\small\hspace{4pt}Cantilever} &
\includegraphics[width=0.18\textwidth]{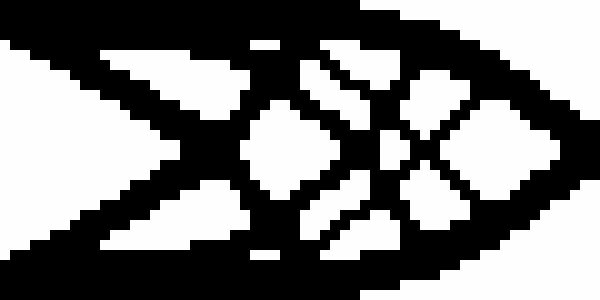} &
\includegraphics[width=0.18\textwidth]{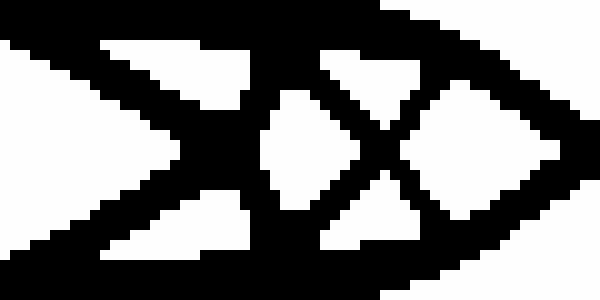} &
\includegraphics[width=0.18\textwidth]{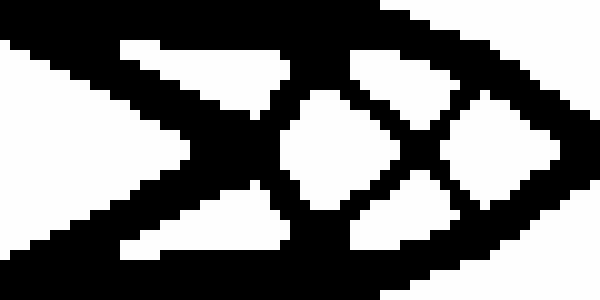} &
\includegraphics[width=0.18\textwidth]{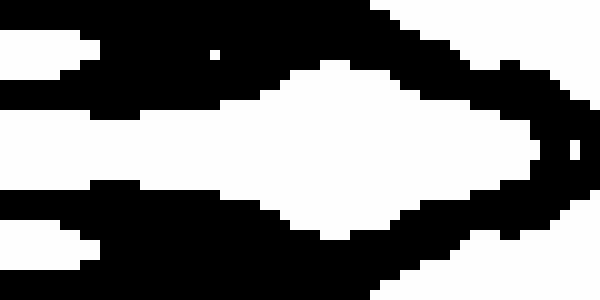} &
\includegraphics[width=0.18\textwidth]{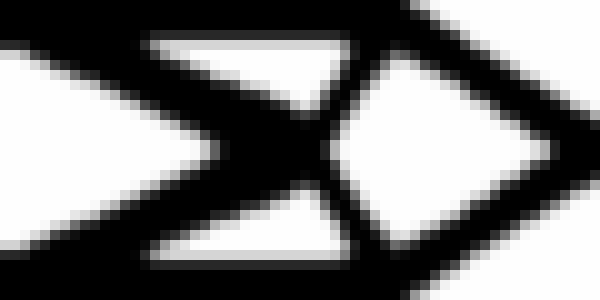} \\[-2pt]
 & \tiny $C=63.69$ & \tiny $C=64.46$ & \tiny $C=63.13$ & \tiny $C=141.17$ & \tiny $C=68.54$ \\[4pt]
\rotatebox{90}{\small\hspace{10pt}MBB} &
\includegraphics[width=0.18\textwidth]{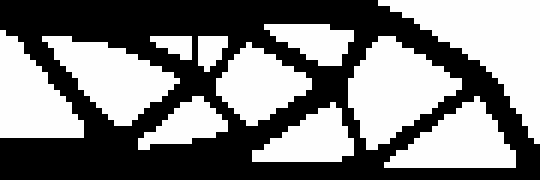} &
\includegraphics[width=0.18\textwidth]{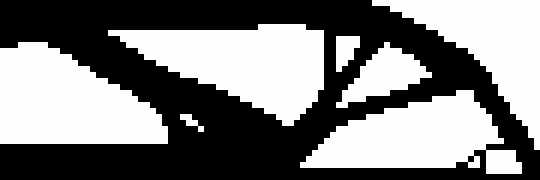} &
\includegraphics[width=0.18\textwidth]{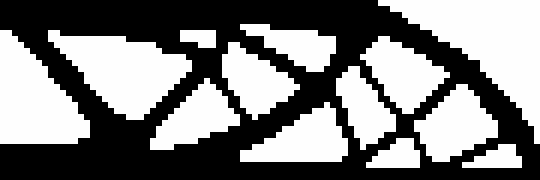} &
\includegraphics[width=0.18\textwidth]{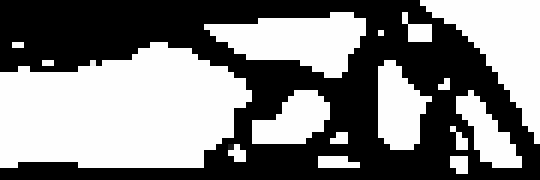} &
\includegraphics[width=0.18\textwidth]{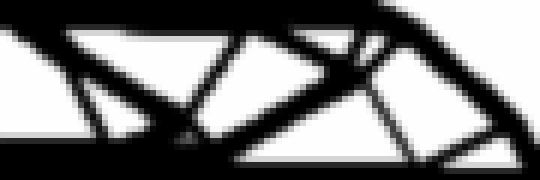} \\[-2pt]
 & \tiny $C=185.16$ & \tiny $C=203.11$ & \tiny $C=190.34$ & \tiny $C=337.66$ & \tiny $C=204.81$ \\[4pt]
\rotatebox{90}{\small\hspace{2pt}Cant.+hole} &
\includegraphics[width=0.18\textwidth]{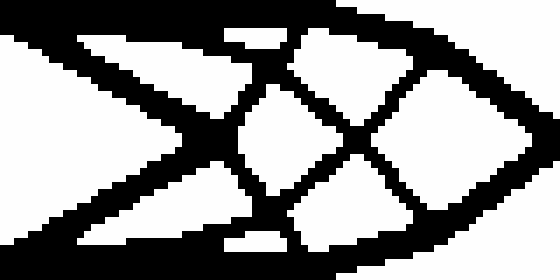} &
\includegraphics[width=0.18\textwidth]{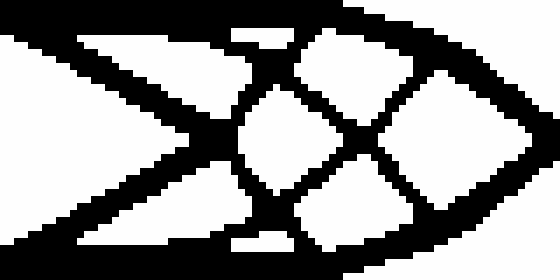} &
\includegraphics[width=0.18\textwidth]{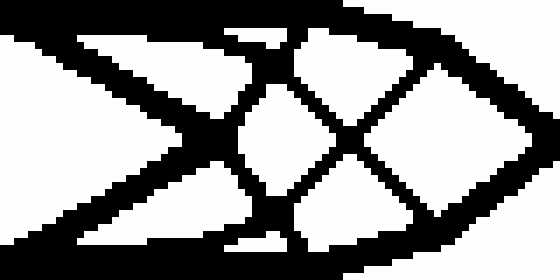} &
\includegraphics[width=0.18\textwidth]{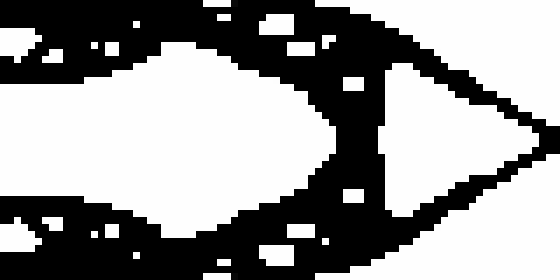} &
\includegraphics[width=0.18\textwidth]{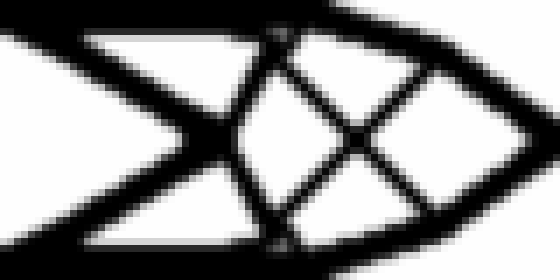} \\[-2pt]
 & \tiny $C=76.24$ & \tiny $C=77.35$ & \tiny $C=76.44$ & \tiny $C=125.58$ & \tiny $C=86.74$ \\[4pt]
\rotatebox{90}{\small\hspace{8pt}Bridge} &
\includegraphics[width=0.18\textwidth]{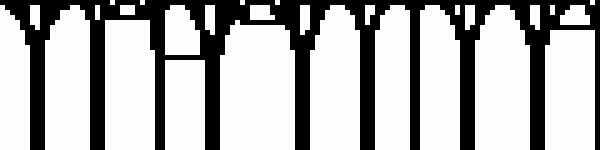} &
\includegraphics[width=0.18\textwidth]{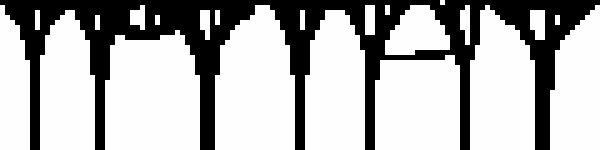} &
\includegraphics[width=0.18\textwidth]{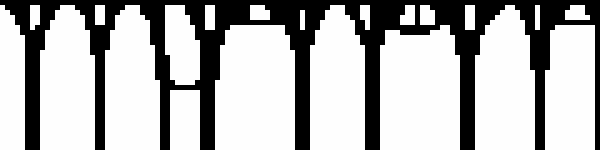} &
\includegraphics[width=0.18\textwidth]{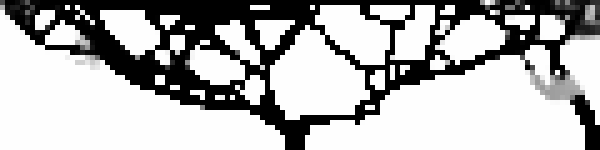} &
\includegraphics[width=0.18\textwidth]{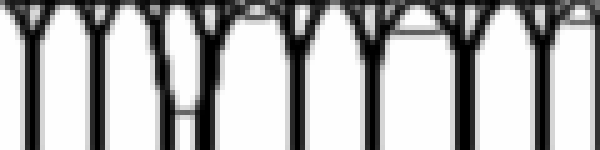} \\[-2pt]
 & \tiny $C=17.95$ & \tiny $C=23.92$ & \tiny $C=19.67$ & \tiny $C=552.55$ & \tiny $C=24.72$ \\[2pt]
\end{tabular}
\caption{Controller comparison (\cref{tab:controllers}): optimized topologies for four
representative problems across five controllers. The LLM, schedule, and
expert controllers produce crisp, structurally similar topologies. The
fixed controller retains gray intermediate densities. The tail-only
controller — which receives no exploration and starts the tail from
uniform density — produces bloated, fragmented designs with
$2$--$30\times$ higher compliance, confirming that the exploration phase is essential. Black = solid, white = void.}
\label{fig:table2}
\end{figure}

\paragraph{Exploration matters.}
The tail-only ablation — which forces $\penal = 1.0$ throughout the main
loop and starts the tail from uniform density — produces designs with
median compliance 211.78, approximately $2.8\times$ worse than any active
controller.
This confirms that the 300-iteration exploration phase is not merely a
warm-up: it discovers structural topologies that the 40-iteration tail
cannot recover from scratch.
Figure~\ref{fig:table2} shows this visually: the tail-only column produces
bloated, fragmented designs with no recognizable truss structure.

\paragraph{The tail matters.}
The fixed controller (no continuation, no tail) achieves only 35.3\% pass
rate with mean grayness 0.169 — confirming that Heaviside projection via
the sharpening tail is essential for producing binary designs.

\begin{figure}[!htbp]
\centering
\includegraphics[width=\textwidth]{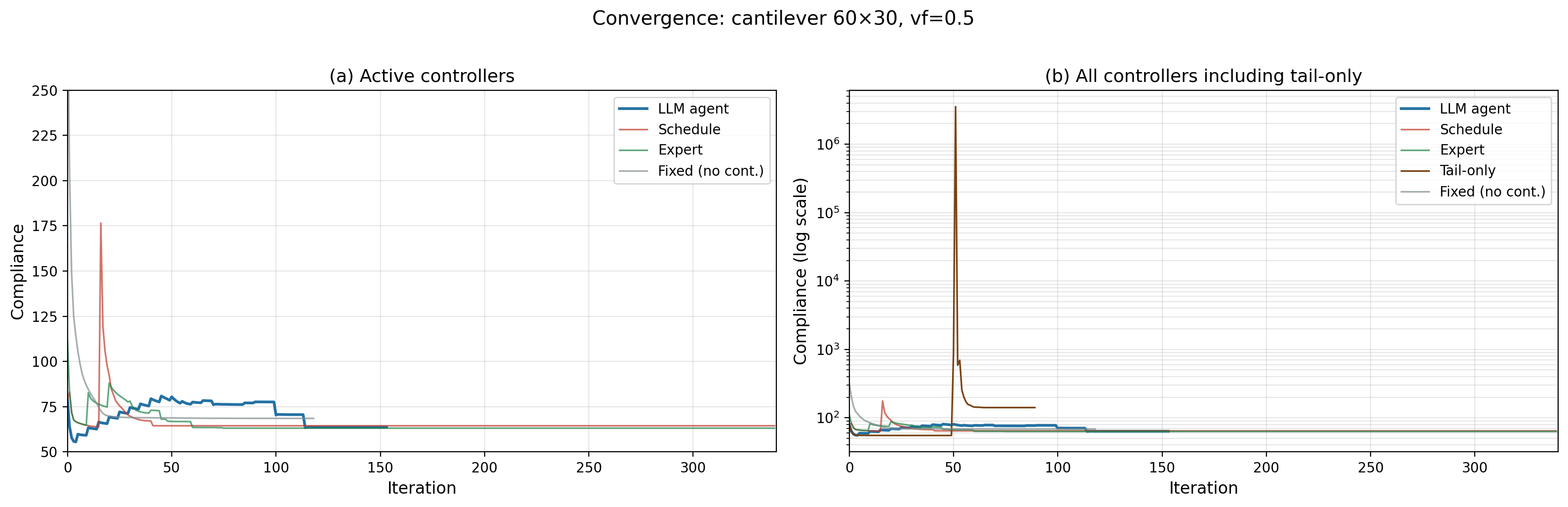}
\caption{Compliance convergence on the cantilever 60$\times$30 benchmark.
(a)~Active controllers on a linear scale: the LLM agent (blue) maintains
higher compliance during the exploration phase (iterations 0--120) before
converging in the tail. The schedule (red) and expert (green) converge
earlier. All active controllers reach similar final compliance ($\approx$63--65).
(b)~All controllers on a log scale: the tail-only controller (brown) spikes
to $3.5 \times 10^6$ during the tail phase as it attempts to build
structure from uniform density, before settling at $C = 141$ —
$2.2\times$ worse than the active controllers.}
\label{fig:convergence}
\end{figure}

\paragraph{Active controllers converge.}
The compliance differences among the four active controllers (LLM, schedule,
expert, three-field) are small: median compliance ranges from 76.24 to 77.35,
a spread of only 1.5\% (Table~\ref{tab:perproblem}).
This suggests that the tail is the primary driver of final topology quality,
with the exploration phase determining the starting point for the tail.
The convergence plot (Fig.~\ref{fig:convergence}) shows that all active
controllers reach similar compliance levels by iteration 150, with
differences emerging primarily in the tail phase.

\subsection{3-D Validation}
\label{sec:results:3d}

Table~\ref{tab:3d} reports results on the two 3-D benchmarks.

\begin{table}[!htbp]
\centering
\caption{3-D controller comparison: 30$\times$15$\times$8 mesh,
$\vf = 0.4$, 300 iterations.}
\label{tab:3d}
\small
\begin{tabular}{lrrrrrr}
\toprule
Problem & LLM & Schedule & Expert & Three-field & Tail-only & Fixed \\
\midrule
cantilever\_3d & \textbf{4.82} & 4.87 & 4.83 & 4.83 & 14.23 & 6.31 \\
mbb\_3d        & \textbf{7.61} & 7.77 & 7.63 & 7.64 & 9.49  & 9.67 \\
\bottomrule
\end{tabular}
\end{table}

\begin{figure}[!htbp]
\centering
\setlength{\tabcolsep}{2pt}
\begin{tabular}{@{}ccccc@{}}
 & \small LLM ($C=4.82$) & \small Schedule ($C=4.87$) & \small Tail-only ($C=14.23$) & \small Fixed ($C=6.31$) \\[2pt]
\rotatebox{90}{\small\hspace{4pt}Isosurface} &
\includegraphics[width=0.22\textwidth]{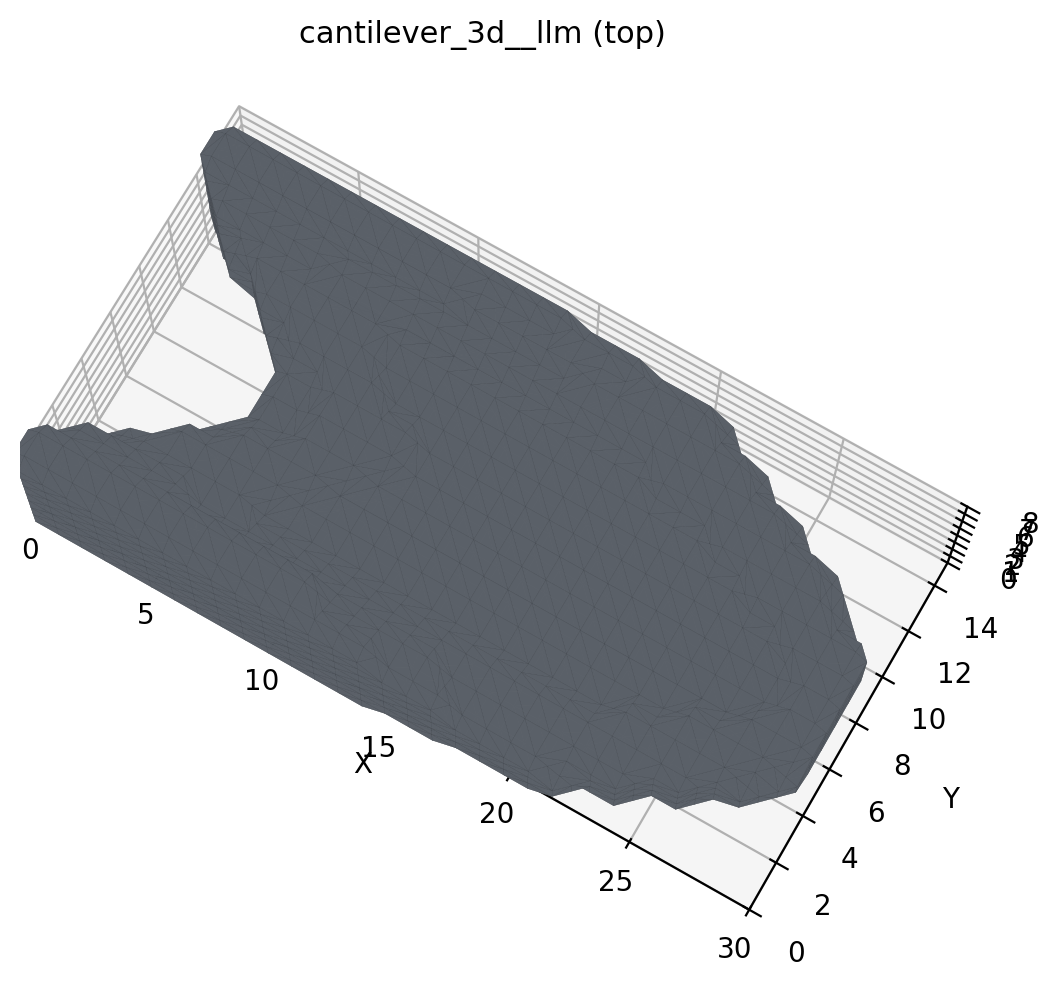} &
\includegraphics[width=0.22\textwidth]{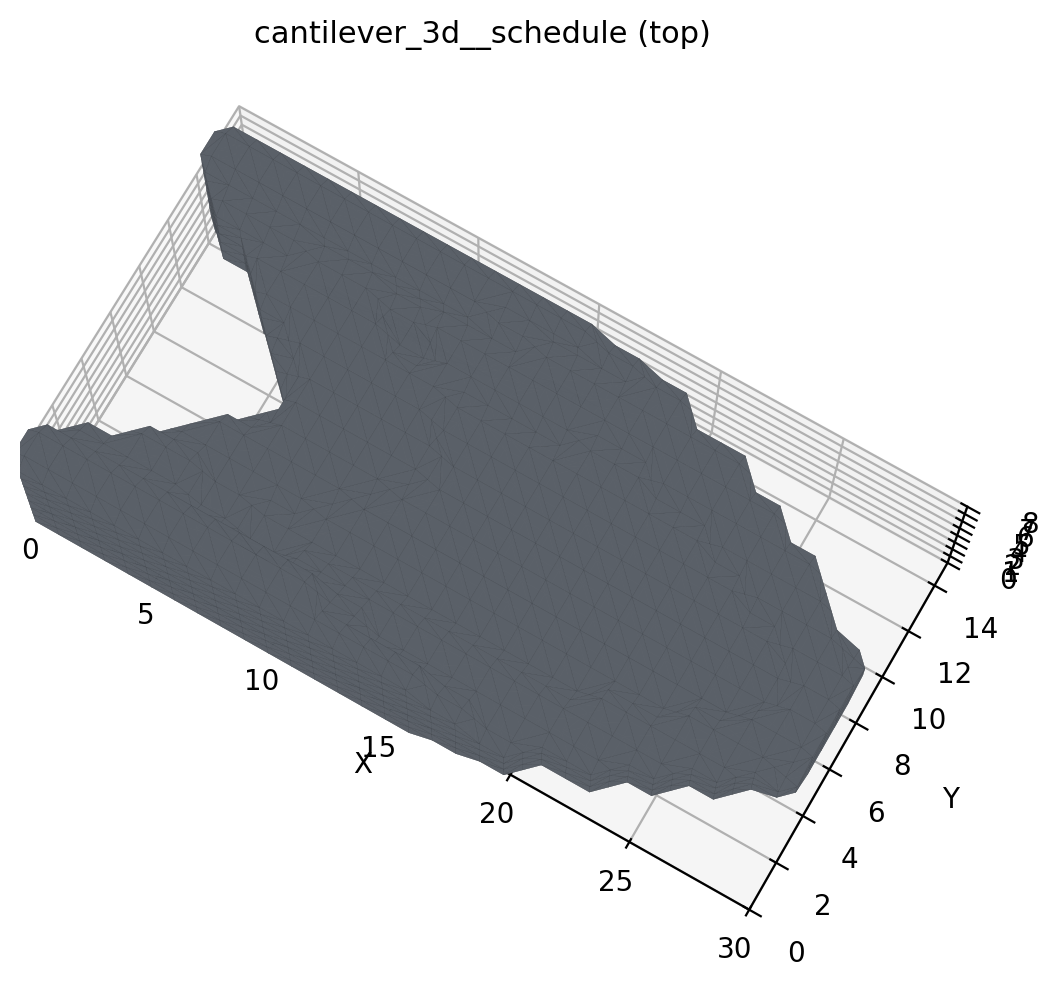} &
\includegraphics[width=0.22\textwidth]{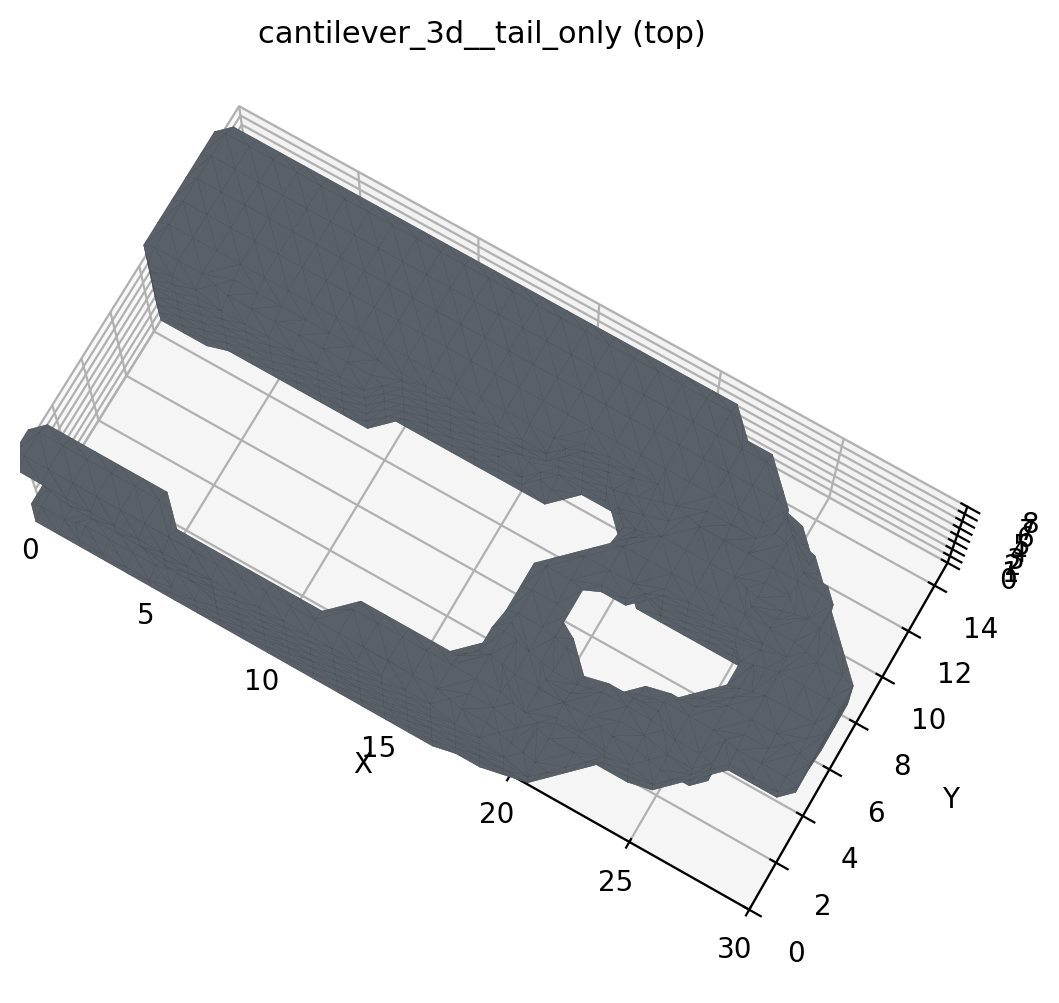} &
\includegraphics[width=0.22\textwidth]{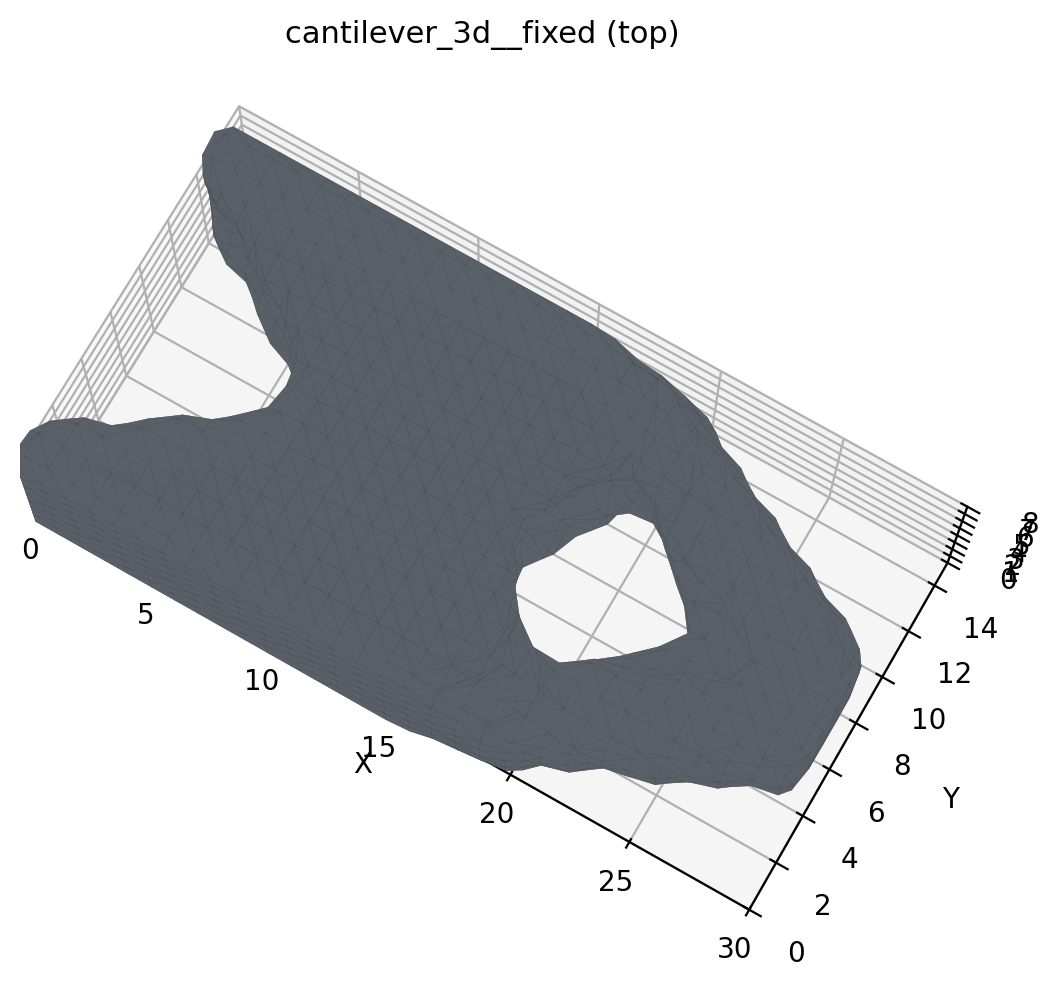} \\[4pt]
\rotatebox{90}{\small\hspace{4pt}X-ray proj.} &
\includegraphics[width=0.22\textwidth]{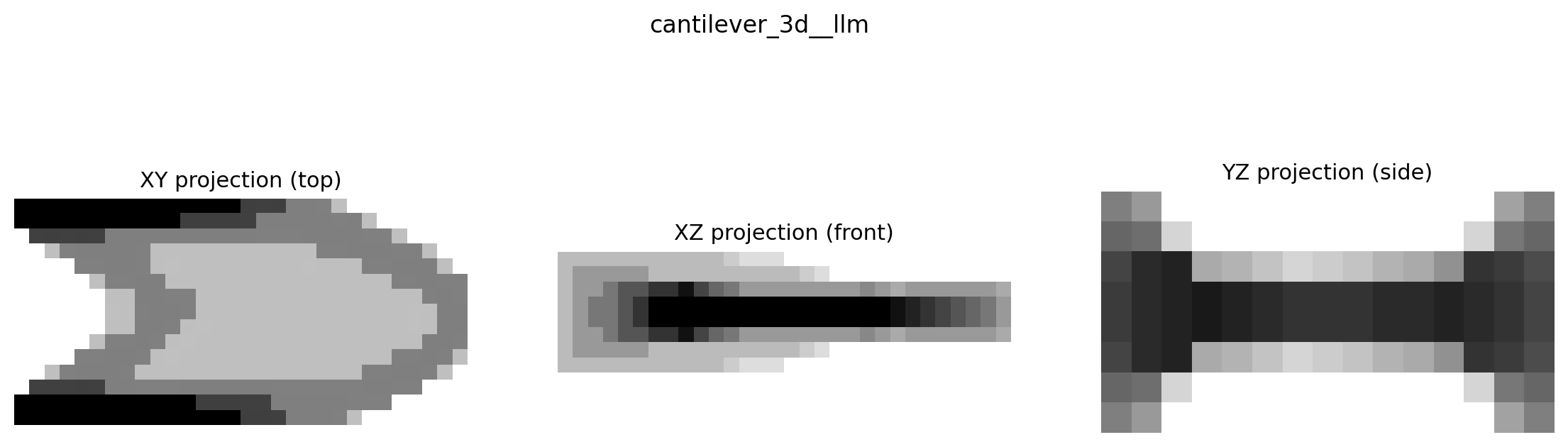} &
\includegraphics[width=0.22\textwidth]{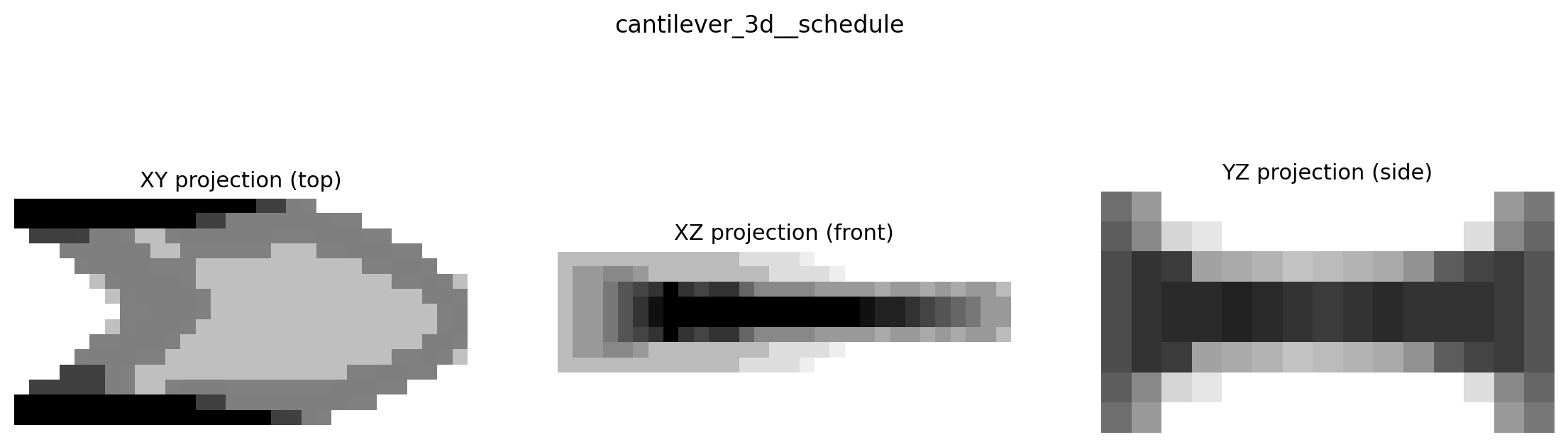} &
\includegraphics[width=0.22\textwidth]{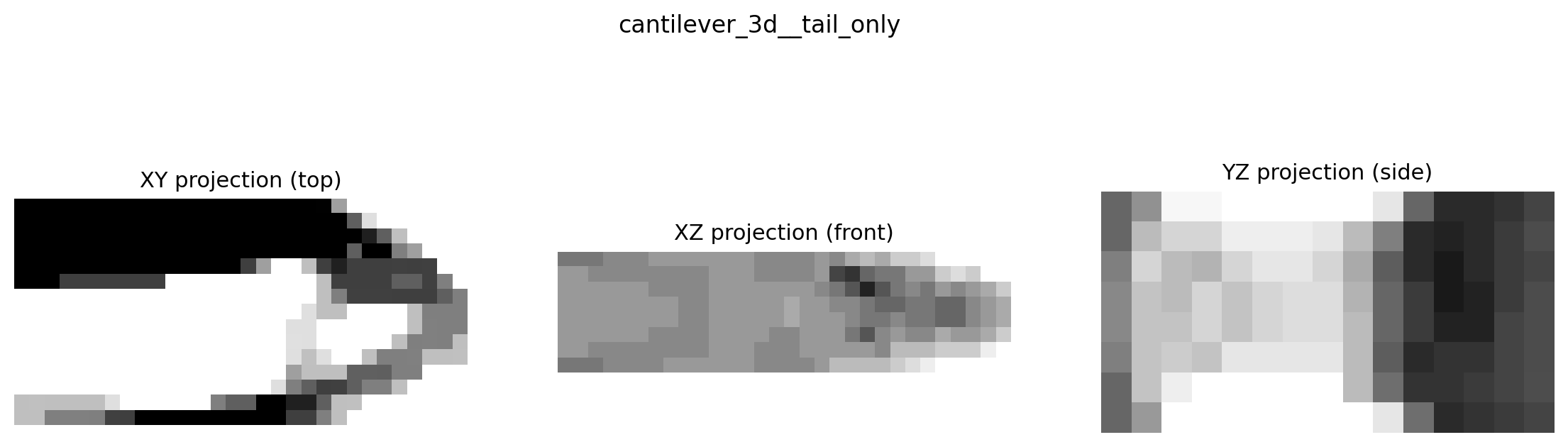} &
\includegraphics[width=0.22\textwidth]{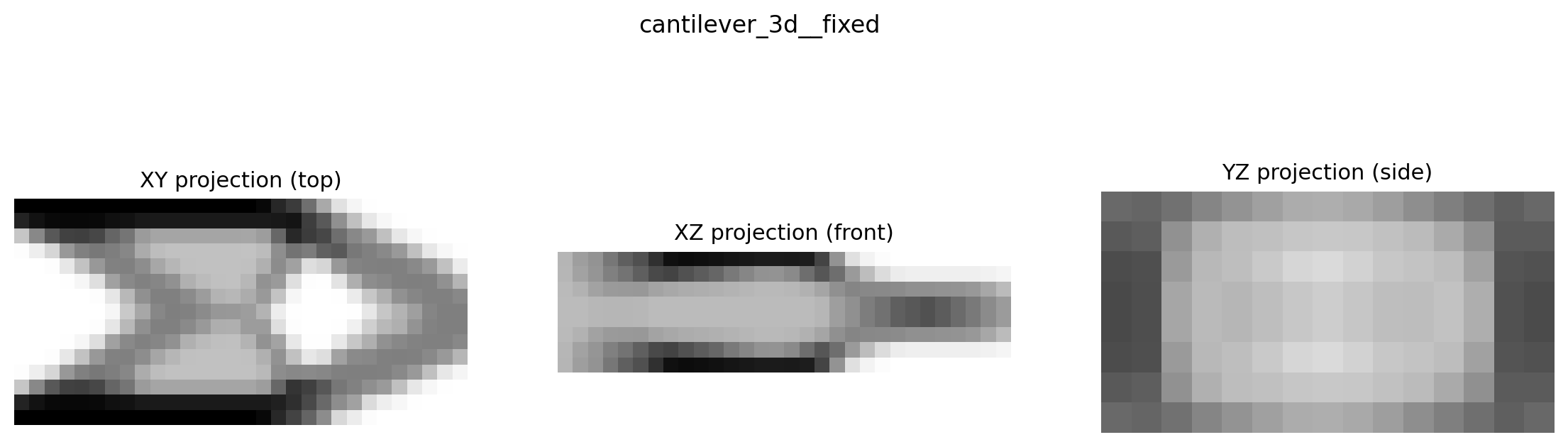} \\[2pt]
\end{tabular}
\caption{3-D cantilever (30$\times$15$\times$8, $\vf=0.4$): isosurface
renders (top, marching cubes at $\rho=0.5$) and X-ray density projections
(bottom, mean along each axis). The LLM and schedule controllers produce
clean wedge-shaped cantilevers with material concentrated at the top and
bottom flanges. The tail-only controller produces a bloated slab with
internal voids — $3\times$ higher compliance. The fixed controller
retains gray intermediate densities visible in the washed-out X-ray
projections.}
\label{fig:3d}
\end{figure}

The \LLM{} controller achieves the lowest compliance on both 3-D problems.
The gap relative to the fixed baseline is substantially larger than in 2-D:
$-23.6\%$ on the 3-D cantilever and $-21.3\%$ on the 3-D \MBB{} beam,
compared to a typical $-7\%$ to $-9\%$ gap in 2-D.
This is consistent with the finding of~\citet{Yang2026LLMController} that the \LLM{}'s
adaptive advantage widens in 3-D, where no established continuation
schedule exists.
The tail-only controller produces designs $2.0$--$3.0\times$ worse than
active controllers, confirming that the exploration ablation result
generalizes to 3-D.
Figure~\ref{fig:3d} shows the isosurface renders and X-ray projections:
the \LLM{} and schedule topologies are visually similar (wedge-shaped
cantilever), while the tail-only and fixed designs are bloated and gray.

\begin{figure}[!htbp]
\centering
\begin{subfigure}[t]{0.48\textwidth}
  \centering
  \includegraphics[width=\textwidth]{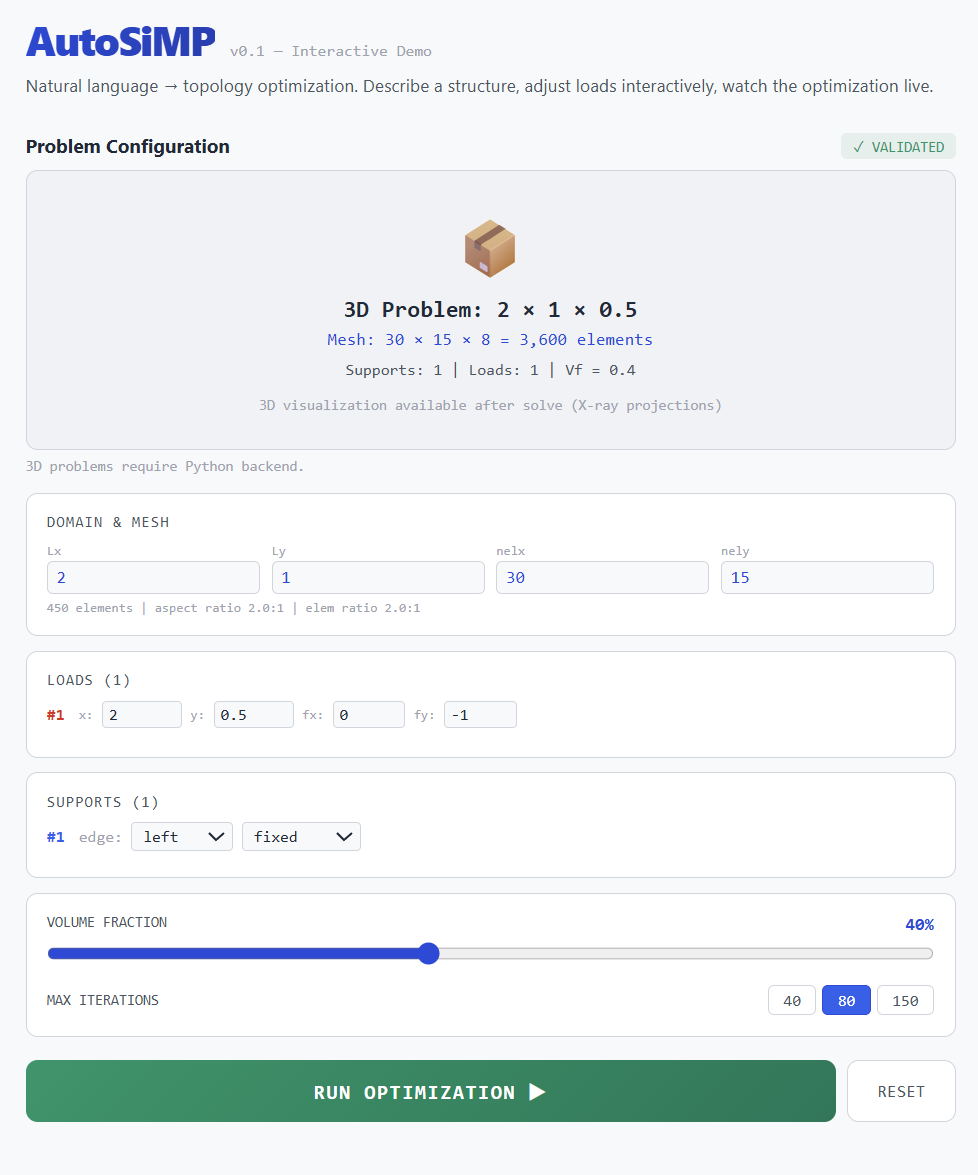}
  \caption{3-D problem configuration in the interactive web demo.
  The validated \PSpec{} shows domain dimensions ($2 \times 1 \times 0.5$),
  mesh resolution ($30 \times 15 \times 8 = 3{,}600$ elements), editable
  load positions, support-type dropdowns, volume-fraction slider, and
  iteration budget selector.
  3-D problems require the Python backend.}
  \label{fig:demo3d:config}
\end{subfigure}\hfill
\begin{subfigure}[t]{0.48\textwidth}
  \centering
  \includegraphics[width=\textwidth]{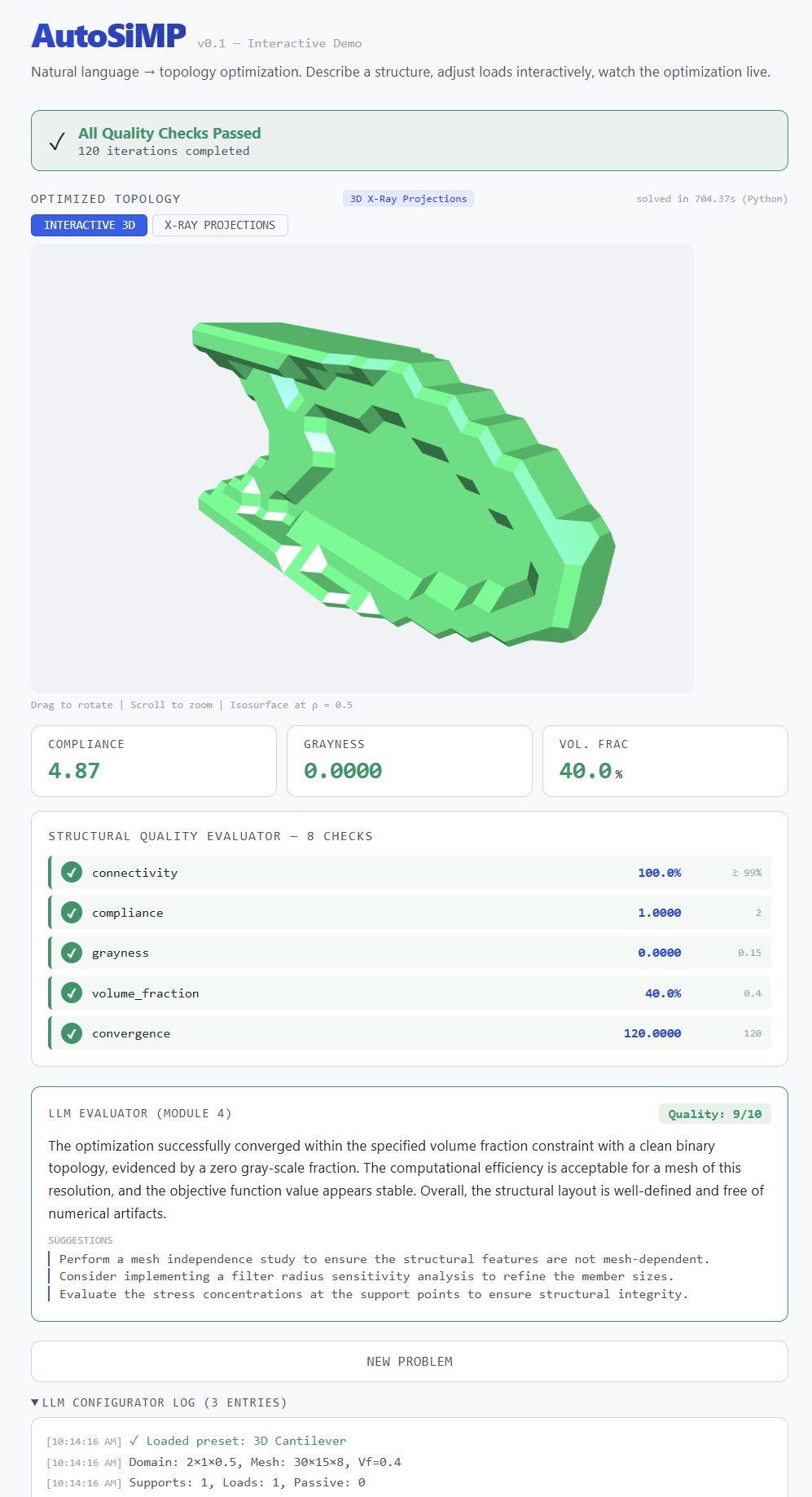}
  \caption{3-D result: interactive isosurface rendering (marching cubes at
  $\rho = 0.5$) with mouse-driven rotation and zoom, compliance and
  grayness metrics, all eight structural quality checks passed, and
  \LLM{} qualitative assessment (quality score 9/10).
  Solved in 704\,s via Python backend.}
  \label{fig:demo3d:result}
\end{subfigure}
\caption{Interactive web demo: 3-D cantilever workflow.
Left: problem configuration with editable domain, mesh, loads, supports,
and solver parameters.
Right: optimized topology with interactive 3-D visualisation, eight-check
evaluator, and \LLM{}-generated qualitative assessment.
This demonstrates the complete autonomous pipeline operating on a 3-D
problem through the browser interface — a capability absent from all
systems in Table~\ref{tab:related}.}
\label{fig:demo3d}
\end{figure}

Figure~\ref{fig:demo3d} shows the complete 3-D workflow through the
interactive web demo: problem configuration (left) and the optimized
result with interactive isosurface inspection, quality evaluation, and
\LLM{} assessment (right).


\subsection{Retry Recovery}
\label{sec:results:table3}

Table~\ref{tab:retry} reports the end-to-end reliability experiment.

\begin{table}[!htbp]
\centering
\caption{Retry recovery: 10 pipeline problems, schedule controller,
300 iterations, \LLM{}-configured specifications.}
\label{tab:retry}
\small
\begin{tabular}{lr}
\toprule
Metric & Value \\
\midrule
Single-shot pass rate   & 10/10 (100\%) \\
With-retries pass rate  & 10/10 (100\%) \\
Mean attempts needed    & 1.0 \\
\bottomrule
\end{tabular}
\end{table}

With the schedule controller, all 10 \LLM{}-configured problems pass all
eight quality checks on the first attempt — no retries are needed.
This result has two implications.
First, the 20\% failure rate observed in \cref{tab:pipeline} (where the \LLM{}
controller was used) is caused entirely by the \LLM{} \emph{controller's}
API-induced instability, not by the \LLM{} \emph{configurator}.
Second, the evaluator and retry mechanism, while not exercised under
deterministic control, provide a safety net for cases where the \LLM{}
controller is used or where the configurator produces edge-case
specifications.

The combination of Tables~1, 2, and 3 supports the central claim: the
configurator is the critical enabler of autonomous operation.
When paired with a deterministic controller, the \LLM{}-configured
pipeline achieves 100\% reliability with near-expert-level compliance
(median $+0.3\%$ penalty).

\FloatBarrier
\section{Discussion}
\label{sec:discussion}


\subsection{Separation of Configuration and Control}

The experimental design deliberately separates the evaluation of the
configurator (\cref{tab:pipeline}) from the controller (\cref{tab:controllers}) and the full loop
(\cref{tab:retry}).
This separation reveals that the two components have complementary
strengths and failure modes.
The configurator is highly reliable (valid specifications on all 10 cases)
but occasionally imprecise (L-bracket load placement).
The \LLM{} controller is high-performing (lowest compliance) but
occasionally unreliable (76.5\% pass rate).
The recommended deployment combines the \LLM{} configurator with the
deterministic schedule controller: this pairing inherits the
configurator's language understanding and the schedule's 100\% reliability.

\subsection{The L-Bracket Outlier}
\label{sec:discussion:lbracket}

The $+119\%$ compliance penalty on the L-bracket warrants discussion.
The configurator correctly identified all structural elements — the domain
is a 2$\times$2 square with the upper-right quadrant voided, the top edge
is fixed, and a horizontal load is applied on the right face.
However, the y-coordinate of the load differed from the ground truth,
producing a different load path and a topology with an additional diagonal
member (Fig.~\ref{fig:table1}).
This is not a \emph{wrong} design in any absolute sense — it is a valid
topology for the specified boundary conditions — but it demonstrates that
the configurator's interpretation of ambiguous spatial descriptions
(``horizontal load at mid-right'') can diverge from the user's intent.
Importantly, this failure mode is \emph{already addressed} by the
interactive web demo (\cref{sec:m_demo}): the preview stage displays
the parsed load position on the domain schematic, allowing the user to
drag it to the intended location before solving.
More broadly, the L-bracket case illustrates a general principle: the
configurator is most reliable for well-known problem archetypes
(cantilever, \MBB{} beam) where the system prompt encodes unambiguous
interpretation rules, and least reliable for problems where spatial
descriptions admit multiple valid interpretations.
For production use, the recommended workflow is therefore
\LLM{}-configure $\to$ visual-verify $\to$ solve, rather than
fully unattended operation.

\subsection{The Tail-Only Fix and Ablation Integrity}

An earlier version of the tail-only controller returned no parameter
updates for all iterations, which caused the solver to use its default
$\penal = 3.0$ — effectively providing a free 300-iteration warm-up.
With the validity gate satisfied from iteration 1, the tail restarted from
an already-converged snapshot, giving tail-only an unfair advantage.
The corrected controller forces $\penal = 1.0$ (below the validity gate)
and the corrected solver resets to uniform density when no valid best exists.
The result: tail-only compliance went from 62.71 (spuriously competitive)
to 141.14 (correctly $2.2\times$ worse) on the standard cantilever,
validating that exploration genuinely matters.


\subsection{Role of the Interactive Web Interface}

The web demo (Section~\ref{sec:m_demo}) serves three purposes.
First, it provides immediate accessibility: a reviewer or practitioner
can verify the end-to-end pipeline without installing Python dependencies
or obtaining solver source code.
Second, it demonstrates features absent from all systems in
Table~\ref{tab:related}: interactive load repositioning, real-time
density visualization during the solve, and rotatable 3-D isosurface
inspection.
Third, the multi-provider \LLM{} configuration (supporting Gemini,
OpenAI, Claude, and custom endpoints) demonstrates that the configurator
architecture is model-agnostic — the system prompt and safety rails,
not the specific \LLM{}, are the critical design elements.
The L-bracket disambiguation problem identified in
\cref{sec:discussion:lbracket} exemplifies how the interactive preview
complements the configurator, turning a potential failure into a
routine visual check.


\subsection{Limitations}

\paragraph{Single-seed results.}
All experiments use seed~42.
The deterministic controllers (schedule, expert, three-field, fixed)
produce bit-identical results across seeds; variance arises only from the
\LLM{} controller's API response variability.
The multi-seed behaviour of the \LLM{} controller was characterized
in~\citet{Yang2026LLMController} across $n=5$ independent seeds, with standard
deviations below $0.05$ on all benchmarks, confirming that the
single-seed results reported here are representative.
Multi-seed validation of the full \AutoSIMP{} pipeline would further
strengthen the statistical claims but would multiply the already
substantial computational budget
($\approx$\,3~hours for 102 2-D runs, $\approx$\,2.7~hours for 12 3-D
runs).

\paragraph{3-D mesh resolution.}
The 3-D benchmarks use a 30$\times$15$\times$8 mesh (3{,}600 elements),
which is coarse by production standards.
Higher-resolution 3-D validation (\eg, 60$\times$30$\times$15) would
strengthen the generalization claim but requires substantially longer
solve times.

\paragraph{LLM model dependency.}
All results use \texttt{gemini-3.1-flash-lite-preview} at temperature~0.
The configurator's behaviour depends on the specific model version;
a model update could alter interpretation rules.
The total API cost for all experiments is under \$1 USD, making
replication trivial.

\paragraph{Scope.}
The current system handles compliance minimization with volume constraints.
Extension to stress-constrained problems, multi-physics formulations, or
manufacturing constraints would require modifications to the system
prompt and evaluator that have not been validated.

\FloatBarrier
\section{Conclusion}
\label{sec:conclusion}

We have presented \AutoSIMP{}, a system that achieves end-to-end
topology optimization from natural language.
The five-module pipeline — configurator, BC generator, \SIMP{} solver,
evaluator, retry loop — transforms a plain-English problem description
into a validated binary topology without domain-expert intervention.

The experimental results establish three findings.
First, the \LLM{} configurator produces valid specifications on all
10 test cases with a median compliance penalty of $+0.3\%$, demonstrating
that frontier language models possess sufficient structural-engineering
knowledge to parse boundary conditions, loads, and passive regions from
informal descriptions.
Second, the controller comparison across 19 benchmarks (17 2-D, 2 3-D)
reveals a quality--reliability trade-off: the \LLM{} controller achieves
the lowest compliance but 76.5\% pass rate, while the deterministic
schedule achieves 100\% pass rate at $+1.5\%$ higher median compliance.
The tail-only ablation confirms that exploration genuinely matters
($2.8\times$ worse without it), and the fixed baseline confirms that the
sharpening tail is essential for binarization.
Third, with the schedule controller, all \LLM{}-configured problems pass
all quality checks on the first attempt, demonstrating that the
\emph{configurator} — not the controller — is the critical enabler of autonomous operation.

The recommended deployment pairs the \LLM{} configurator with the
deterministic schedule controller, combining natural-language accessibility
with 100\% reliability at near-expert compliance.
An interactive web demo with browser-side solving, interactive load
editing, and 3-D isosurface visualization is provided to enable
immediate verification and exploration of the pipeline without
installation (Fig.~\ref{fig:demo}).
The complete codebase and demo will be made publicly available upon
acceptance. Among the systems compared in Table~\ref{tab:related},
\AutoSIMP{} is the only one that combines natural-language input,
boundary-condition generation, a topology optimization solver,
structural quality evaluation, retry recovery, and an interactive
visual interface.

\section*{Code availability}

The complete source code for reproducing all results in this paper —
including the three-field \SIMP{} solver, \LLM{} configurator agent,
boundary-condition generator, structural evaluator, all baseline
controllers, the experiment runner, all 19 benchmark specifications,
and the interactive web demo (React frontend + Flask backend) — will be
made publicly available upon journal acceptance.

\bibliographystyle{unsrtnat}
\bibliography{ref}

@article{Bendsoe1989,
  author    = {Bends{\o}e, Martin P.},
  title     = {Optimal shape design as a material distribution problem},
  journal   = {Structural Optimization},
  year      = {1989},
  volume    = {1},
  number    = {4},
  pages     = {193--202},
  doi       = {10.1007/BF01650949}
}

@article{Sigmund2001,
  author    = {Sigmund, Ole},
  title     = {A 99 line topology optimization code written in {Matlab}},
  journal   = {Structural and Multidisciplinary Optimization},
  year      = {2001},
  volume    = {21},
  number    = {2},
  pages     = {120--127},
  doi       = {10.1007/s001580050176}
}

@article{Andreassen2011,
  author    = {Andreassen, Erik and Clausen, Anders and Schevenels, Mattias
               and Lazarov, Boyan S. and Sigmund, Ole},
  title     = {Efficient topology optimization in {MATLAB} using 88 lines of
               code},
  journal   = {Structural and Multidisciplinary Optimization},
  year      = {2011},
  volume    = {43},
  number    = {1},
  pages     = {1--16},
  doi       = {10.1007/s00158-010-0594-7}
}

@article{Sigmund2013,
  author    = {Sigmund, Ole and Maute, Kurt},
  title     = {Topology optimization approaches},
  journal   = {Structural and Multidisciplinary Optimization},
  year      = {2013},
  volume    = {48},
  number    = {6},
  pages     = {1031--1055},
  doi       = {10.1007/s00158-013-0978-6}
}

@article{Deaton2014,
  author    = {Deaton, Joshua D. and Grandhi, Ramana V.},
  title     = {A survey of structural and multidisciplinary continuum topology
               optimization: post 2000},
  journal   = {Structural and Multidisciplinary Optimization},
  year      = {2014},
  volume    = {49},
  number    = {1},
  pages     = {1--38},
  doi       = {10.1007/s00158-013-0956-z}
}

@article{Wang2011,
  author    = {Wang, Fengwen and Lazarov, Boyan S. and Sigmund, Ole},
  title     = {On projection methods, convergence and robust formulations
               in topology optimization},
  journal   = {Structural and Multidisciplinary Optimization},
  year      = {2011},
  volume    = {43},
  number    = {6},
  pages     = {767--784},
  doi       = {10.1007/s00158-010-0602-y}
}

@article{Lazarov2016,
  author    = {Lazarov, Boyan S. and Wang, Fengwen and Sigmund, Ole},
  title     = {Length scale and manufacturability in density-based topology
               optimization},
  journal   = {Archive of Applied Mechanics},
  year      = {2016},
  volume    = {86},
  number    = {1},
  pages     = {189--218},
  doi       = {10.1007/s00419-015-1106-4}
}

@article{Bourdin2001,
  author    = {Bourdin, Blaise},
  title     = {Filters in topology optimization},
  journal   = {International Journal for Numerical Methods in Engineering},
  year      = {2001},
  volume    = {50},
  number    = {9},
  pages     = {2143--2158},
  doi       = {10.1002/nme.116}
}

@article{Guest2004,
  author    = {Guest, James K. and Pr{\'{e}}vost, Jean H. and Belytschko, Ted},
  title     = {Achieving minimum length scale in topology optimization using
               nodal design variables and projection functions},
  journal   = {International Journal for Numerical Methods in Engineering},
  year      = {2004},
  volume    = {61},
  number    = {2},
  pages     = {238--254},
  doi       = {10.1002/nme.1064}
}

@article{Aage2017,
  author    = {Aage, Niels and Andreassen, Erik and Lazarov, Boyan S.
               and Sigmund, Ole},
  title     = {Giga-voxel computational morphogenesis for structural design},
  journal   = {Nature},
  year      = {2017},
  volume    = {550},
  number    = {7674},
  pages     = {84--86},
  doi       = {10.1038/nature23911}
}

@article{Sosnovik2019,
  author    = {Sosnovik, Ivan and Oseledets, Ivan},
  title     = {Neural networks for topology optimization},
  journal   = {Russian Journal of Numerical Analysis and Mathematical
               Modelling},
  year      = {2019},
  volume    = {34},
  number    = {4},
  pages     = {215--223},
  doi       = {10.1515/rnam-2019-0018}
}

@article{Yu2019,
  author    = {Yu, Yonggyun and Hur, Taeil and Jung, Jaeho and Jang, In Gwun},
  title     = {Deep learning for determining a near-optimal topological
               design without any iteration},
  journal   = {Structural and Multidisciplinary Optimization},
  year      = {2019},
  volume    = {59},
  number    = {3},
  pages     = {787--799},
  doi       = {10.1007/s00158-018-2101-5}
}

@article{Cang2019,
  author    = {Cang, Ruijin and Yao, Hope and Ren, Yi},
  title     = {One-shot generation of near-optimal topology through
               theory-driven machine learning},
  journal   = {Computer-Aided Design},
  year      = {2019},
  volume    = {109},
  pages     = {12--21},
  doi       = {10.1016/j.cad.2018.12.008}
}

@article{Abueidda2020,
  author    = {Abueidda, Diab W. and Koric, Seid and Sobh, Nahed A.},
  title     = {Topology optimization of 2{D} structures with nonlinearities
               using deep learning},
  journal   = {Computers \& Structures},
  year      = {2020},
  volume    = {237},
  pages     = {106283},
  doi       = {10.1016/j.compstruc.2020.106283}
}

@inproceedings{Wei2022,
  author    = {Wei, Jason and Wang, Xuezhi and Schuurmans, Dale
               and Bosma, Maarten and Ichter, Brian and Xia, Fei
               and Chi, Ed and Le, Quoc and Zhou, Denny},
  title     = {Chain-of-thought prompting elicits reasoning in large
               language models},
  booktitle = {Advances in Neural Information Processing Systems},
  year      = {2022},
  volume    = {35},
  pages     = {24824--24837}
}

@inproceedings{Yao2023ReAct,
  author    = {Yao, Shunyu and Zhao, Jeffrey and Yu, Dian and Du, Nan
               and Shafran, Izhak and Narasimhan, Karthik and Cao, Yuan},
  title     = {{ReAct}: Synergizing reasoning and acting in language models},
  booktitle = {International Conference on Learning Representations},
  year      = {2023}
}

@inproceedings{Park2023,
  author    = {Park, Joon Sung and O'Brien, Joseph C. and Cai, Carrie J.
               and Morris, Meredith Ringel and Liang, Percy
               and Bernstein, Michael S.},
  title     = {Generative agents: Interactive simulacra of human behavior},
  booktitle = {ACM Symposium on User Interface Software and Technology},
  year      = {2023},
  pages     = {1--22},
  doi       = {10.1145/3586183.3606763}
}

@inproceedings{Driess2023PaLME,
  author    = {Driess, Danny and Xia, Fei and Sajjadi, Mehdi S. M.
               and Lynch, Corey and Chowdhery, Aakanksha and Ichter, Brian
               and Wahid, Ayzaan and Tompson, Jonathan and Vuong, Quan
               and Yu, Tianhe and Huang, Wenlong and Chebotar, Yevgen
               and Sermanet, Pierre and Duckworth, Daniel and Levine, Sergey
               and Vanhoucke, Vincent and Hausman, Karol
               and Toussaint, Marc and Greff, Klaus and Zeng, Andy
               and Mordatch, Igor and Florence, Pete},
  title     = {{PaLM-E}: An embodied multimodal language model},
  booktitle = {International Conference on Machine Learning},
  year      = {2023},
  pages     = {8469--8488}
}

@techreport{Chen2021Codex,
  author      = {Chen, Mark and Tworek, Jerry and Jun, Heewoo
                 and Yuan, Qiming and de Oliveira Pinto, Henrique Pond{\'{e}}
                 and Kaplan, Jared and Edwards, Harrison and Burda, Yuri
                 and Joseph, Nicholas and Brockman, Greg and Ray, Alex
                 and Puri, Raul and Krueger, Gretchen and Petrov, Michael
                 and Khlaaf, Heidy and Sastry, Girish and Mishkin, Pamela
                 and Chan, Brooke and Gray, Scott and Ryder, Nick
                 and Pavlov, Mikhail and Power, Alethea and Kaiser, Lukasz
                 and Bavarian, Mohammad and Winter, Clemens
                 and Tillet, Philippe and Such, Felipe Petroski
                 and Cummings, Dave and Plappert, Matthias
                 and Chantzis, Fotios and Barnes, Elizabeth
                 and Herbert-Voss, Ariel and Guss, William Hebgen
                 and Nichol, Alex and Paino, Alex and Tezak, Nikolas
                 and Tang, Jie and Babuschkin, Igor and Balaji, Suchir
                 and Jain, Shantanu and Saunders, William
                 and Hesse, Christopher and Carr, Andrew N. and Leike, Jan
                 and Achiam, Josh and Misra, Vedant and Morikawa, Evan
                 and Radford, Alec and Knight, Matthew and Brundage, Miles
                 and Murati, Mira and Mayer, Katie and Welinder, Peter
                 and McGrew, Bob and Amodei, Dario and McCandlish, Sam
                 and Sutskever, Ilya and Zaremba, Wojciech},
  title       = {Evaluating large language models trained on code},
  institution = {OpenAI},
  year        = {2021},
  number      = {arXiv:2107.03374},
  note        = {arXiv preprint}
}

@article{Boiko2023,
  author    = {Boiko, Daniil A. and MacKnight, Robert and Kline, Ben
               and Gomes, Gabe},
  title     = {Autonomous chemical research with large language models},
  journal   = {Nature},
  year      = {2023},
  volume    = {624},
  pages     = {570--578},
  doi       = {10.1038/s41586-023-06792-0}
}

@inproceedings{Liu2024LLMOptim,
  title={Large language models as evolutionary optimizers},
  author={Liu, Shengcai and Chen, Caishun and Qu, Xinghua and Tang, Ke and Ong, Yew-Soon},
  booktitle={2024 IEEE Congress on Evolutionary Computation (CEC)},
  pages={1--8},
  year={2024},
  organization={IEEE},
  doi={10.1109/CEC60901.2024.10611913}
}

@inproceedings{Yang2024OPRO,
  author    = {Yang, Chengrun and Wang, Xuezhi and Lu, Yifeng
               and Liu, Hanxiao and Le, Quoc V. and Zhou, Denny
               and Chen, Xinyun},
  title     = {Large language models as optimizers},
  booktitle = {International Conference on Learning Representations},
  year      = {2024}
}

@article{Romera2024FunSearch,
  title={Mathematical discoveries from program search with large language models},
  author={Romera-Paredes, Bernardino and Barekatain, Mohammadamin and Novikov, Alexander
          and Balog, Matej and Kumar, M. Pawan and Dupont, Emilien
          and Ruiz, Francisco J. R. and Ellenberg, Jordan S. and Wang, Pengming
          and Fawzi, Omar and Kohli, Pushmeet and Fawzi, Alhussein},
  journal={Nature},
  volume={625},
  number={7995},
  pages={468--475},
  year={2024},
  doi={10.1038/s41586-023-06924-6}
}

@article{Bell2023PyAMG,
  title={PyAMG: algebraic multigrid solvers in python},
  author={Bell, Nathan and Olson, Luke N and Schroder, Jacob and Southworth, Ben},
  journal={Journal of Open Source Software},
  volume={8},
  number={87},
  pages={5495},
  year={2023},
  doi={10.21105/joss.05495}
}

@article{BendsoeKikuchi1988,
  author  = {Bends{\o}e, Martin P. and Kikuchi, Noboru},
  title   = {Generating optimal topologies in structural design using
             a homogenization based method},
  journal = {Computer Methods in Applied Mechanics and Engineering},
  year    = {1988},
  volume  = {71},
  number  = {2},
  pages   = {197--224},
  doi     = {10.1016/0045-7825(88)90086-2}
}

@article{ZhouRozvany1991,
  author  = {Zhou, M. and Rozvany, G. I. N.},
  title   = {The {COC} algorithm, {Part II}: Topological, geometrical
             and generalized shape optimization},
  journal = {Computer Methods in Applied Mechanics and Engineering},
  year    = {1991},
  volume  = {89},
  number  = {1--3},
  pages   = {309--336},
  doi     = {10.1016/0045-7825(91)90046-9}
}

@article{BendsoeSigmund1999,
  author  = {Bends{\o}e, Martin P. and Sigmund, Ole},
  title   = {Material interpolation schemes in topology optimization},
  journal = {Archive of Applied Mechanics},
  year    = {1999},
  volume  = {69},
  number  = {9--10},
  pages   = {635--654},
  doi     = {10.1007/s004190050248}
}

@article{BrunsTortorelli2001,
  author  = {Bruns, T. E. and Tortorelli, D. A.},
  title   = {Topology optimization of non-linear elastic structures
             and compliant mechanisms},
  journal = {Computer Methods in Applied Mechanics and Engineering},
  year    = {2001},
  volume  = {190},
  number  = {26--27},
  pages   = {3443--3459},
  doi     = {10.1016/S0045-7825(00)00278-4}
}

@article{StoMeSvanberg2001,
  author  = {Stolpe, Mathias and Svanberg, Krister},
  title   = {On the trajectories of penalization methods for topology
             optimization},
  journal = {Structural and Multidisciplinary Optimization},
  year    = {2001},
  volume  = {21},
  number  = {2},
  pages   = {128--139},
  doi     = {10.1007/s001580050177}
}

@article{RojasStolpe2015,
  author  = {Rojas-Labanda, Susana and Stolpe, Mathias},
  title   = {Automatic penalty continuation in structural topology
             optimization},
  journal = {Structural and Multidisciplinary Optimization},
  year    = {2015},
  volume  = {52},
  number  = {6},
  pages   = {1205--1221},
  doi     = {10.1007/s00158-015-1277-1}
}

@article{FerrariSigmund2020,
  author  = {Ferrari, Federico and Sigmund, Ole},
  title   = {A new generation 99 line {Matlab} code for compliance
             topology optimization and its extension to {3D}},
  journal = {Structural and Multidisciplinary Optimization},
  year    = {2020},
  volume  = {62},
  pages   = {2211--2228},
  doi     = {10.1007/s00158-020-02629-w}
}

@article{AutoProjTO2025,
  title={Automatic projection parameter increase for three-field density-based topology optimization},
  author={Dunning, Peter and Wein, Fabian},
  journal={Structural and Multidisciplinary Optimization},
  volume={68},
  number={2},
  pages={33},
  year={2025},
  doi={10.1007/s00158-025-03968-2}
}

@article{Woldseth2022,
  author  = {Woldseth, Rebekka V. and Aage, Niels and
             B{\ae}rentzen, Jakob Andreas and Sigmund, Ole},
  title   = {On the use of artificial neural networks in topology
             optimisation},
  journal = {Structural and Multidisciplinary Optimization},
  year    = {2022},
  volume  = {65},
  number  = {10},
  pages   = {294},
  doi     = {10.1007/s00158-022-03347-1}
}

@article{Shin2023review,
  title={Topology optimization via machine learning and deep learning: a review},
  author={Shin, Seungyeon and Shin, Dongju and Kang, Namwoo},
  journal={Journal of Computational Design and Engineering},
  volume={10},
  number={4},
  pages={1736--1766},
  year={2023},
  doi={10.1093/jcde/qwad072}
}

@article{Kallioras2020,
  author  = {Kallioras, Nikolaos Ath. and Kazakis, Georgios and
             Lagaros, Nikos D.},
  title   = {Accelerated topology optimization by means of deep
             learning},
  journal = {Structural and Multidisciplinary Optimization},
  year    = {2020},
  volume  = {62},
  number  = {3},
  pages   = {1185--1212},
  doi     = {10.1007/s00158-020-02545-z}
}

@techreport{Hoyer2019,
  author      = {Hoyer, Stephan and Sohl-Dickstein, Jascha and
                 Greydanus, Sam},
  title       = {Neural reparameterization improves structural
                 optimization},
  institution = {arXiv},
  year        = {2019},
  number      = {arXiv:1909.04240},
  note        = {arXiv preprint}
}

@article{Chandrasekhar2021TOuNN,
  author  = {Chandrasekhar, Aaditya and Suresh, Krishnan},
  title   = {{TOuNN}: Topology optimization using neural networks},
  journal = {Structural and Multidisciplinary Optimization},
  year    = {2021},
  volume  = {63},
  number  = {3},
  pages   = {1135--1149},
  doi     = {10.1007/s00158-020-02748-4}
}

@article{NieTopologyGAN2021,
  author  = {Nie, Zhenguo and Lin, Tong and Jiang, Haoliang and
             Kara, Levent B.},
  title   = {{TopologyGAN}: Topology optimization using generative
             adversarial networks based on physical fields over the
             initial domain},
  journal = {ASME Journal of Mechanical Design},
  year    = {2021},
  volume  = {143},
  number  = {3},
  pages   = {031715},
  doi     = {10.1115/1.4049533}
}

@article{MazeDiffusion2023,
  author  = {Maz{\'{e}}, Fran{\c{c}}ois and Ahmed, Faez},
  title   = {Diffusion models beat {GANs} on topology optimization},
  journal = {Proceedings of the AAAI Conference on Artificial
             Intelligence},
  year    = {2023},
  volume  = {37},
  number  = {8},
  pages   = {9108--9116},
  doi     = {10.1609/aaai.v37i8.26093}
}

@article{DengSOLO2022,
  title={Self-directed online machine learning for topology optimization},
  author={Deng, Changyu and Wang, Yizhou and Qin, Can and Fu, Yun and Lu, Wei},
  journal={Nature Communications},
  volume={13},
  number={1},
  pages={388},
  year={2022},
  doi={10.1038/s41467-021-27713-7}
}

@article{BrownRL2022,
  author  = {Brown, Nathan K. and Garland, Anthony P. and
             Fadel, Georges M. and Li, Gang},
  title   = {Deep reinforcement learning for engineering design through
             topology optimization of elementally discretized design
             domains},
  journal = {Materials \& Design},
  year    = {2022},
  volume  = {218},
  pages   = {110672},
  doi     = {10.1016/j.matdes.2022.110672}
}

@inproceedings{BiedenkappDAC2020,
  author    = {Biedenkapp, Andr{\'{e}} and Bozkurt, H. Furkan and
               Eimer, Theresa and Hutter, Frank and Lindauer, Marius},
  title     = {Dynamic Algorithm Configuration: Foundation of a New
               Meta-Algorithmic Framework},
  booktitle = {Proceedings of the 24th European Conference on
               Artificial Intelligence (ECAI 2020)},
  year      = {2020},
  pages     = {427--434},
  doi       = {10.3233/FAIA200122}
}

@article{AdriaensenAutoDAC2022,
  author  = {Adriaensen, Steven and Biedenkapp, Andr{\'{e}} and
             Shala, Gresa and Awad, Noor and Eimer, Theresa and
             Lindauer, Marius and Hutter, Frank},
  title   = {Automated Dynamic Algorithm Configuration},
  journal = {Journal of Artificial Intelligence Research},
  year    = {2022},
  volume  = {75},
  pages   = {1633--1699},
  doi     = {10.1613/jair.1.13922}
}

@inproceedings{SchickToolformer2023,
  author    = {Schick, Timo and Dwivedi-Yu, Jane and
               Dess{\`{\i}}, Roberto and Raileanu, Roberta and
               Lomeli, Maria and Zettlemoyer, Luke and
               Cancedda, Nicola and Scialom, Thomas},
  title     = {Toolformer: Language Models Can Teach Themselves to
               Use Tools},
  booktitle = {Advances in Neural Information Processing Systems},
  year      = {2023},
  pages    = {2510--2544},
  volume    = {36}
}

@article{WangVoyager2024,
  author    = {Wang, Guanzhi and Xie, Yuqi and Jiang, Yunfan and
               Mandlekar, Ajay and Xiao, Chaowei and Zhu, Yuke and
               Fan, Linxi and Anandkumar, Anima},
  title     = {Voyager: An Open-Ended Embodied Agent with Large
               Language Models},
  journal   = {Transactions on Machine Learning Research},
  year      = {2024}
}

@inproceedings{MadaanSelfRefine2023,
  author    = {Madaan, Aman and Tandon, Niket and Gupta, Prakhar and
               Hallinan, Skyler and Gao, Luyu and Wiegreffe, Sarah and
               Alon, Uri and Dziri, Nouha and Prabhumoye, Shrimai and
               Yang, Yiming and Gupta, Shailendra and
               Majumder, Bodhisattwa Prasad and Hermann, Katherine and
               Welleck, Sean and Yazdanbakhsh, Amir and Clark, Peter},
  title     = {Self-{Refine}: Iterative Refinement with Self-Feedback},
  booktitle = {Advances in Neural Information Processing Systems},
  year      = {2023},
  pages     = {1127--1148},
  volume    = {36}
}

@inproceedings{ShinnReflexion2023,
  author    = {Shinn, Noah and Cassano, Federico and
               Gopinath, Ashwin and Narasimhan, Karthik and
               Yao, Shunyu},
  title     = {Reflexion: Language Agents with Verbal Reinforcement
               Learning},
  booktitle = {Advances in Neural Information Processing Systems},
  year      = {2023},
  pages     = {8634--8652},
  volume    = {36}
}

@inproceedings{YeReEvo2024,
  author    = {Ye, Haoran and Wang, Jiarui and Cao, Zhiguang and
               Berto, Federico and Hua, Chuanbo and Kim, Haeyeon and
               Park, Jinkyoo and Song, Guojie},
  title     = {{ReEvo}: Large Language Models as Hyper-Heuristics with
               Reflective Evolution},
  booktitle = {Advances in Neural Information Processing Systems},
  year      = {2024}
}

@article{WuEvoLLM2024,
  title={Evolutionary computation in the era of large language model: Survey and roadmap},
  author={Wu, Xingyu and Wu, Sheng-hao and Wu, Jibin and Feng, Liang and Tan, Kay Chen},
  journal={IEEE Transactions on Evolutionary Computation},
  volume={29},
  number={2},
  pages={534--554},
  year={2025},
  doi={10.1109/TEVC.2024.3506731}
}

@article{Hayashi2020,
  author  = {Hayashi, Kazuki and Ohsaki, Makoto},
  title   = {Reinforcement learning and graph embedding for binary truss
             topology optimization under stress and displacement
             constraints},
  journal = {Frontiers in Built Environment},
  year    = {2020},
  volume  = {6},
  pages   = {59},
  doi     = {10.3389/fbuil.2020.00059}
}

@inproceedings{Rios2023,
  author    = {Rios, Thiago and Menzel, Stefan and Sendhoff, Bernhard},
  title     = {Large Language and Text-to-{3D} Models for Engineering
               Design Optimization},
  booktitle = {2023 {IEEE} Symposium Series on Computational Intelligence
               ({SSCI})},
  year      = {2023},
  pages     = {1704--1711},
  doi       = {10.1109/SSCI52147.2023.10371898}
}

@misc{Yang2026LLMController,
  author    = {Yang, Shaoliang and Wang, Jun and Wang, Yunsheng},
  title     = {Large Language Models as Optimization Controllers: Adaptive
               Continuation for {SIMP} Topology Optimization},
  year      = {2026},
  eprint    = {2603.25099},
  archivePrefix = {arXiv},
  primaryClass  = {cs.CE},
  url       = {https://arxiv.org/abs/2603.25099}
}

@article{LLMShapeOpt2024,
  author    = {Zhang, Xinxin and Xu, Zhuoqun and Zhu, Guangpu and
               Tay, Chien Ming Jonathan and Cui, Yongdong and
               Khoo, Boo Cheong and Zhu, Lailai},
  title     = {Using large language models for parametric shape optimization},
  journal   = {Physics of Fluids},
  year      = {2025},
  volume    = {37},
  number    = {8},
  pages     = {083601},
  doi       = {10.1063/5.0273363}
}

@article{LMTO2025,
  author    = {Liang, Zelong and Zhang, Yuan-Fang and Wang, Yingjun and Li, Weihua},
  title     = {Integrating large models with topology optimization for
               conceptual design realization},
  journal   = {Advanced Engineering Informatics},
  year      = {2025},
  volume    = {67},
  pages     = {103524},
  doi       = {10.1016/j.aei.2025.103524}
}

@techreport{LLMStructAnalysis2025,
  author      = {Liang, Haoran and Talebi Kalaleh, Mohammad and Mei, Qipei},
  title       = {Integrating Large Language Models for Automated Structural
                 Analysis},
  institution = {arXiv},
  year        = {2025},
  number      = {arXiv:2504.09754},
  note        = {arXiv preprint}
}

@inproceedings{LLMOPT2024,
  author    = {Jiang, Caigao and Shu, Xiang and Qian, Hong and Lu, Xingyu
               and Zhou, Jun and Zhou, Aimin and Yu, Yang},
  title     = {{LLMOPT}: Learning to Define and Solve General Optimization
               Problems from Scratch},
  booktitle = {International Conference on Learning Representations},
  year      = {2025}
}

@techreport{Banga2018,
  author      = {Banga, Saurabh and Gehani, Harsh and Bhilare, Sanket and
                 Patel, Sagar and Kara, Levent Burak},
  title       = {{3D} topology optimization using convolutional neural networks},
  institution = {arXiv},
  year        = {2018},
  number      = {arXiv:1808.07440},
  note        = {arXiv preprint}
}

@article{Sun2020,
  author    = {Sun, Hongbo and Ma, Ling},
  title     = {Generative design by using exploration approaches of
               reinforcement learning in density-based structural topology
               optimization},
  journal   = {Designs},
  year      = {2020},
  volume    = {4},
  number    = {2},
  pages     = {10},
  doi       = {10.3390/designs4020010}
}

@article{Jang2022,
  author    = {Jang, Seungki and Yoo, Sungmin and Kang, Namwoo},
  title     = {Generative design by reinforcement learning: enhancing the
               diversity of topology optimization designs},
  journal   = {Computer-Aided Design},
  year      = {2022},
  volume    = {146},
  pages     = {103225},
  doi       = {10.1016/j.cad.2022.103225}
}

@techreport{TransformerTO2025,
  author      = {Lutheran, Aaron and Das, Srijan and Tabarraei, Alireza},
  title       = {Transformer-based Topology Optimization},
  institution = {arXiv},
  year        = {2025},
  number      = {arXiv:2509.05800},
  note        = {arXiv preprint}
}

@techreport{FeaGPT2025,
  author      = {Qi, Yupeng and Xu, Ran and Chu, Xu},
  title       = {{FeaGPT}: an End-to-End Agentic-{AI} for Finite Element Analysis},
  institution = {arXiv},
  year        = {2025},
  number      = {arXiv:2510.21993},
  note        = {arXiv preprint}
}

@article{MCPSIM2026,
  author    = {Park, Donggeun and Moon, Hyeonbin and Ryu, Seunghwa},
  title     = {A self-correcting multi-agent {LLM} framework for language-based
               physics simulation and explanation},
  journal   = {npj Artificial Intelligence},
  year      = {2026},
  volume    = {2},
  pages     = {10},
  doi       = {10.1038/s44387-025-00057-z}
}

@article{LLMFEAEval2025,
  author    = {Shafiq, Ossama and Rahmat, Amin and Alexiadis, Alessio and
               Ghiassi, Bahman},
  title     = {Evaluating the Performance of Large Language Models for Geometry
               and Simulation File Generation in Physics-Based Simulations},
  journal   = {Applied Sciences},
  year      = {2025},
  volume    = {15},
  number    = {22},
  pages     = {12114},
  doi       = {10.3390/app152212114}
}

\end{document}